\documentclass[aps,prl,nofootinbib,preprintnumbers,amsmath,amssymb,latexsym,array,enumerate,letter,twocolumn,superscriptaddress]{revtex4}
\pdfoutput=1
\usepackage{amssymb}
\usepackage{amsmath}
\usepackage{epsfig}
\usepackage{hyperref}
\usepackage{breakurl}
\usepackage{xcolor}

\makeatletter
\def\simgt{\mathrel{\lower2.5pt\vbox{\lineskip=0pt\baselineskip=0pt
           \hbox{$>$}\hbox{$\sim$}}}}
\def\simlt{\mathrel{\lower2.5pt\vbox{\lineskip=0pt\baselineskip=0pt
           \hbox{$<$}\hbox{$\sim$}}}}
\makeatother

\def\be{\begin{equation}}
\def\ee{\end{equation}}
\newcommand{\bea}{\begin{eqnarray}}
\newcommand{\eea}{\end{eqnarray}}

\newcommand{\pv}{\langle\phi\rangle}

\begin{document}

\leftline{LCTP-18-01}

\title{Exotic Sterile Neutrinos and Pseudo-Goldstone Phenomenology}

\author{Bibhushan Shakya}
\affiliation{Department of Physics, University of Cincinnati,
Cincinnati, OH 45221, USA}
\affiliation{Leinweber Center for Theoretical
Physics, University of Michigan, Ann Arbor, MI 48109}
\author{James D. Wells}
\affiliation{Leinweber Center for Theoretical
Physics, University of Michigan, Ann Arbor, MI 48109}

\begin{abstract}

We study the phenomenology of a light (GeV scale) sterile neutrino sector and the pseudo-Goldstone boson (not the majoron) associated with a global symmetry in this sector that is broken at a high scale. Such scenarios can be motivated from considerations of singlet fermions from a hidden sector coupling to active neutrinos via heavy right-handed seesaw neutrinos, effectively giving rise to a secondary, low-energy seesaw framework. This framework involves rich phenomenology with observable implications for cosmology, dark matter, and direct searches, involving novel sterile neutrino dark matter production mechanisms from the pseudo-Goldstone-mediated scattering or decay, modifications of BBN bounds on sterile neutrinos, suppression of canonical sterile neutrino decay channels at direct search experiments, late injection of an additional population of neutrinos in the Universe after neutrino decoupling, and measurable dark radiation.
  
\end{abstract}

\maketitle

\section{Motivation}
\label{sec:motivation}

The most straightforward explanation of tiny neutrino masses is the seesaw mechanism, involving Standard Model (SM)  singlet (sterile) right-handed neutrinos at a heavier scale. GUT (grand unified theory) scale seesaw models \cite{Minkowski:1977sc, Mohapatra:1979ia,Mohapatra:1980yp, Yanagida:1980xy, GellMann:1980vs, Schechter:1980gr} accomplish this with $\mathcal{O}(1)$ couplings with heavy sterile neutrinos at  $M\sim 10^{10}-10^{15}$ GeV. However, the seesaw mechanism is also consistent with masses below the electroweak scale, which are motivated by connections to dark matter (DM) and leptogenesis as in the neutrino Minimal Standard Model ($\nu$MSM) \cite{Asaka:2005an,Asaka:2005pn,Asaka:2006nq} and involve potentially rich phenomenology in cosmology, indirect detection, and direct searches \cite{Abazajian:2012ys,Adhikari:2016bei}. 

Drastic departures from the seesaw phenomenology is possible if additional symmetries or particles exist in the sterile neutrino sector beyond the basic elements of the seesaw framework (see e.g. \cite{Ma:2017dbl,Alanne:2017sip,Sayre:2005yh}). Since the Majorana mass of a pure singlet fermion is expected to lie at the ultraviolet (UV) cutoff scale of the theory (such as the GUT or Planck scale), light sterile neutrinos are plausibly charged under some symmetry. If this symmetry is related to lepton number, the sterile neutrino masses emerge from a low scale of lepton number breaking \cite{Chikashige:1980ui,Gelmini:1980re, Georgi:1981pg, Schechter:1981cv, Gelmini:1982rr,Lindner:2011it,Escudero:2016tzx}; rich phenomenology ensues from the existence of additional scalars \cite{Maiezza:2015lza,Nemevsek:2016enw,Maiezza:2016ybz,Dev:2017dui} and massive gauge bosons \cite{Mohapatra:1979ia,Mohapatra:1980yp,Keung:1983uu} or a (pseudo-) Goldstone boson, the majoron \cite{Chikashige:1980ui,Gelmini:1980re, Georgi:1981pg, Schechter:1981cv, Gelmini:1982rr,Lindner:2011it,Escudero:2016tzx}.   

This symmetry can, however, be confined entirely to the sterile neutrino sector. This can occur, for instance, if the sterile neutrinos originate from a separate hidden sector. As discussed in the next section, even with a GUT-scale realization of the seesaw mechanism, exotic fermions from hidden sectors that couple to the GUT scale right-handed neutrinos develop couplings to the SM neutrinos, mimicking a low energy seesaw setup, effectively acting as light sterile neutrinos akin to those studied in, e.g.\,the $\nu$MSM.

In this letter, we consider a global symmetry confined to, and spontaneously broken in, such a light (GeV scale) exotic sterile neutrino sector, and study the phenomenology of the pseudo-Goldstone boson $\eta$ of this broken symmetry. GeV scale sterile neutrinos can equilibrate with the thermal bath and dominate the energy density of the Universe before big bang nucleosynthesis (BBN) \cite{Asaka:2006ek} --- their interplay with $\eta$ can therefore give rise to novel cosmological scenarios. The $\eta$ phenomenology can be very different from the more familiar majoron phenomenology, as the scale of symmetry breaking, lepton number breaking, and sterile neutrino masses are all different, which can enable several new possibilities for cosmology, dark matter, and direct searches that are not possible in the majoron framework.

\section{Charged-Singlet Seesaws}

The canonical seesaw mechanism involves three SM-singlet, right-handed neutrinos $N_i$, with: 
\be
\mathcal{L}\supset y_{ij} L_i h N_j+M_i \bar{N}^c_i N_i.
\label{eq:lagrangian1}
\ee
$L_i$ and $h$ are the SM lepton doublet and Higgs fields, and $y_{ij}$ are dimensionless Yukawa couplings. The hierarchy $M\gg y v$ (where $v$ is the Higgs vacuum expectation value (vev)) leads to the familiar seesaw mechanism, resulting in active and sterile neutrino masses $m_a\sim y^2 v^2/M,~ m_s\sim M$, with an active-sterile mixing angle sin\,$\theta\sim y\,v/M$. $M\sim 10^{14}$ GeV produces the desired neutrino masses for $y\sim\mathcal{O}(1)$, whereas $M\sim$ GeV requires $y\sim10^{-7}$. 

A global or gauged $U(1)_{\text{lepton}}$ or $U(1)_{B-L}$ symmetry for $N_i$ \cite{Chikashige:1980ui,Gelmini:1980re, Georgi:1981pg, Schechter:1981cv, Gelmini:1982rr,Lindner:2011it,Escudero:2016tzx} precludes the Majorana mass term; the lagrangian is instead
\be
\mathcal{L}\supset y_{ij} L_i h N_j+x_i \phi \bar{N}^c_i N_i + \lambda (H^\dagger H) \phi^2 + V(\phi).
\label{eq:lagrangian2}
\ee
A vev for the exotic Higgs field $\phi$, appropriately charged under the lepton or $B-L$ symmetry, breaks the symmetry and produces sterile neutrino masses $M_i\sim x \pv$. If the symmetry is global, a physical light degree of freedom, the Goldstone boson, known as the majoron, emerges \cite{Chikashige:1980ui,Gelmini:1980re}. 

In this paper, we consider instead a global symmetry, for instance a $U(1)'$, that is confined to the sterile neutrinos and does not extend to any SM field. Such a symmetry forbids both terms in Eq.\,\ref{eq:lagrangian1}. However, a scalar field $\phi$ carrying the opposite $U(1)'$ charge to $N_i$ enables the higher dimensional operator $ \frac{1}{\Lambda} L h N \phi$, where $\Lambda$ is a UV-cutoff scale.\,\footnote{Such operators have been studied in the context of supersymmetry \cite{Cleaver:1997nj,Langacker:1998ut,ArkaniHamed:2000bq,ArkaniHamed:2000kj,Wells:2004di}, including the freeze-in production of sterile neutrino DM \cite{Roland:2014vba,Roland:2015yoa,Roland:2016gli}.} A $\phi$ vev breaks the $U(1)'$ and produces the Yukawa interaction term from Eq.\,\ref{eq:lagrangian1} with the effective Yukawa coupling $y\sim \lambda_1\pv/\Lambda$; thus such an operator also provides a natural explanation for the tiny Yukawas in terms of the hierarchy between the two scales $\pv$ and $\Lambda$. Next, we discuss a UV completion of this setup in terms of singlet fermions from a hidden sector that couple to heavy right-handed seesaw neutrinos.

\subsection{``Sterile neutrinos" from a hidden sector with a heavy right-handed neutrino portal}

We start with the original seesaw motivation of pure singlet, heavy (scale $M$, possibly close to the GUT scale) right-handed neutrinos that couple to SM neutrinos through Yukawa terms $y_{ij}L_ihN_j$. If the $N_j$ also act as portals to a hidden sector\,\footnote{For recent studies of right-handed neutrinos acting as portals to a hidden/dark sector, see \cite{Falkowski:2009yz,Falkowski:2011xh,Pospelov:2011ha,Pospelov:2012gm,Cherry:2014xra,Berryman:2017twh,Batell:2017cmf,Schmaltz:2017oov}.}, this invokes the generic prospect of an analogous Yukawa term $y'_{ij}L'_ih'N_j$, where $L'_ih'$ is a singlet combination of hidden sector fields analogous to $L_i h$. Integrating out the $N_i$ produces the following dimension-5 operators connecting the visible and hidden sectors \footnote{We assume that the $N_i$ sector is sufficiently extended and general that one cannot rotate the $L, L'$ system to suppress couplings of any particular $L, L'$ to the $N_i$ sector. } :
\be
\mathcal{L}\supset \frac{1}{M}y^2 (L h)^2 +\frac{1}{M}y y' (L h)(L'h')+\frac{1}{M}y'^2 (L'h')^2.
\label{eq:lagrangian4}
\ee
In the above we have ignored flavor structure and dropped indices for simplicity, assuming all $y_{ij} (y_{ij})$ are roughly the same, so that the above terms should only be taken as approximate. If the hidden sector scalar acquires a vev $v'$, the above can be rewritten as
\be
\mathcal{L}\supset  \frac{1}{\Lambda_{\rm eff}} (L h)^2 + y_{\text{eff}}L h L'+ M_{\text{eff}} L' L'
\label{eq:lagrangian5}
\ee
where we have defined $\Lambda_{\rm eff}^{-1}\equiv y^2/M$, $y_{\text{eff}}\equiv y y' v'/M$, and $M_{\text{eff}}\equiv y'^2 v'^2/M$. Here, the first term accounts for the active neutrino masses $y^2 v^2/M$ from the primary seesaw involving integrating out the pure singlet neutrinos $N_i$. The latter two terms give a similar contribution to the active neutrino masses from the secondary seesaw resulting from integrating out the $L'_i$ fermions (note the analogy between Eq.\,\ref{eq:lagrangian5} and Eq.\,\ref{eq:lagrangian1}).

The mixing angle between the active neutrinos and these hidden sector singlets $L'$ is approximately
\be
\sin\theta'\sim\frac{y_{\text{eff}}\,v}{M_{\text{eff}}}=\frac{y v}{y' v'}=\sqrt{\frac{m_a}{M_{\text{eff}}}}\,,
\ee
which is the relation expected from a seesaw framework. Therefore, light sterile neutrinos that appear to satisfy the seesaw relation could have exotic origins in a hidden sector connected via a high scale neutrino portal, with symmetries unrelated to the SM, and themselves obtain light masses via the seesaw mechanism.\,\footnote{This setup holds similarities with extended seesaw models \cite{Chun:1995js,Ma:1995gf,Zhang:2011vh,Barry:2011wb,Boulebnane:2017fxw}, which also employ a seesaw suppression for sterile neutrino masses to naturally accommodate an eV scale sterile neutrino.} We will henceforth ignore the integrated out ``true" right-handed seesaw neutrinos and work with the effective field theory (EFT) in Eq.\,\ref{eq:lagrangian5}, switching the notation $N_i$ to refer to these light sterile states $L'$, whose phenomenology we will pursue in this paper.

\subsection{Pseudo-Goldstone Boson}

The spontaneous breaking of the global $U(1)'$ by $\pv\equiv f$ gives rise to a massless Goldstone boson, which we will call the $\eta$-boson. It is conjectured that non-perturbative gravitational effects explicitly break global symmetries, leading to a pseudo-Goldstone boson mass of order $m_\eta^2\sim f^3/M_{Pl}$ via an operator of the form $\frac{\phi^5}{M_{Pl}}$ \cite{Rothstein:1992rh,Akhmedov:1992hi}.\,\footnote{An explicit $U(1)'$ breaking Goldstone mass term is also possible. A small $\eta$ mass is also generated from the Yukawa coupling \cite{Frigerio:2011in}, but is negligible for the parameters we are interested in.} For generality, we treat $m_\eta$ as a free parameter, but this approximate mass scale should be kept in mind. 

Next, we draw the distinction between the $\eta$-boson and the more familiar majoron \cite{Chikashige:1980ui,Gelmini:1980re, Georgi:1981pg, Schechter:1981cv, Gelmini:1982rr,Lindner:2011it,Escudero:2016tzx}. For both, couplings to (both active and sterile) neutrinos are proportional to the neutrino mass suppressed by the scale of symmetry breaking, as expected for Goldstone bosons, hence several phenomenological bounds on the majoron symmetry breaking scale \cite{Gelmini:1982zz,Gelmini:1982rr,Jungman:1991sh,Pilaftsis:1993af,Garcia-Cely:2017oco,Adhikari:2016bei} are also applicable to $\eta$. However, the majoron is associated with the breaking of lepton number --- a symmetry shared by the SM leptons as well as the sterile neutrinos --- and the sterile neutrino mass scale approximately coincides with the scale of lepton number breaking. This results in the majoron being much lighter that the sterile neutrinos. Furthermore, this scaling leads to specific relations between majoron couplings and sterile neutrino masses, which drives many of the constraints on majorons \cite{Gelmini:1982zz,Gelmini:1982rr,Jungman:1991sh,Pilaftsis:1993af,Garcia-Cely:2017oco,Adhikari:2016bei}. 

In contrast, these energy scales are distinct in the $\eta$ framework: the symmetry breaking scale $f$ (i.e., the scale of $U(1)'$ breaking) is independent of the breaking of lepton number (at the much higher real seesaw scale $M$) and is also distinct from the sterile neutrino mass scale ($M_{\rm eff}\sim f^2/M$), which, as discussed above, is suppressed by a seesaw mechanism. The ability to vary them independently opens up phenomenologically interesting regions of parameter space. Furthermore, the sterile neutrino masses $M_{\rm eff}\sim f^2/M$ can be comparable to the $\eta$-boson mass $m_\eta^2\sim f^3/{M_{\rm Pl}}$ (if $f\sim M^2/M_{\rm Pl}$); this coincidence of mass scales can carry important implications for cosmology and DM, as we will see later. 

\section{Framework and Phenomenology}

We focus on the low-energy effective theory containing three sterile neutrinos (which we have reset to the label $N_i$ rather than $L'$), and the pseudo-Goldstone boson $\eta$. We treat $m_{N_i}, f$, and $m_\eta$ as independent parameters. We assume $m_{N_i}\sim$ GeV scale, and $y_{ij}$ are correspondingly small in a natural way that matches the measured $\Delta m_{\nu}^2$ and mixings among the light active neutrinos. We will consider the interesting and widely studied possibility that the lightest sterile neutrino $N_1$ is DM, which is especially appealing given recent claims of a 3.5 keV X-ray line \cite{Bulbul:2014sua,Boyarsky:2014jta} compatible with decays of a 7 keV sterile neutrino DM particle. We also assume $f\gg v$; the $U(1)'$ breaking singlet scalar is then decoupled and irrelevant for phenomenology.

\begin{figure}[t!]
\includegraphics[width=3.1in, height=1.9in]{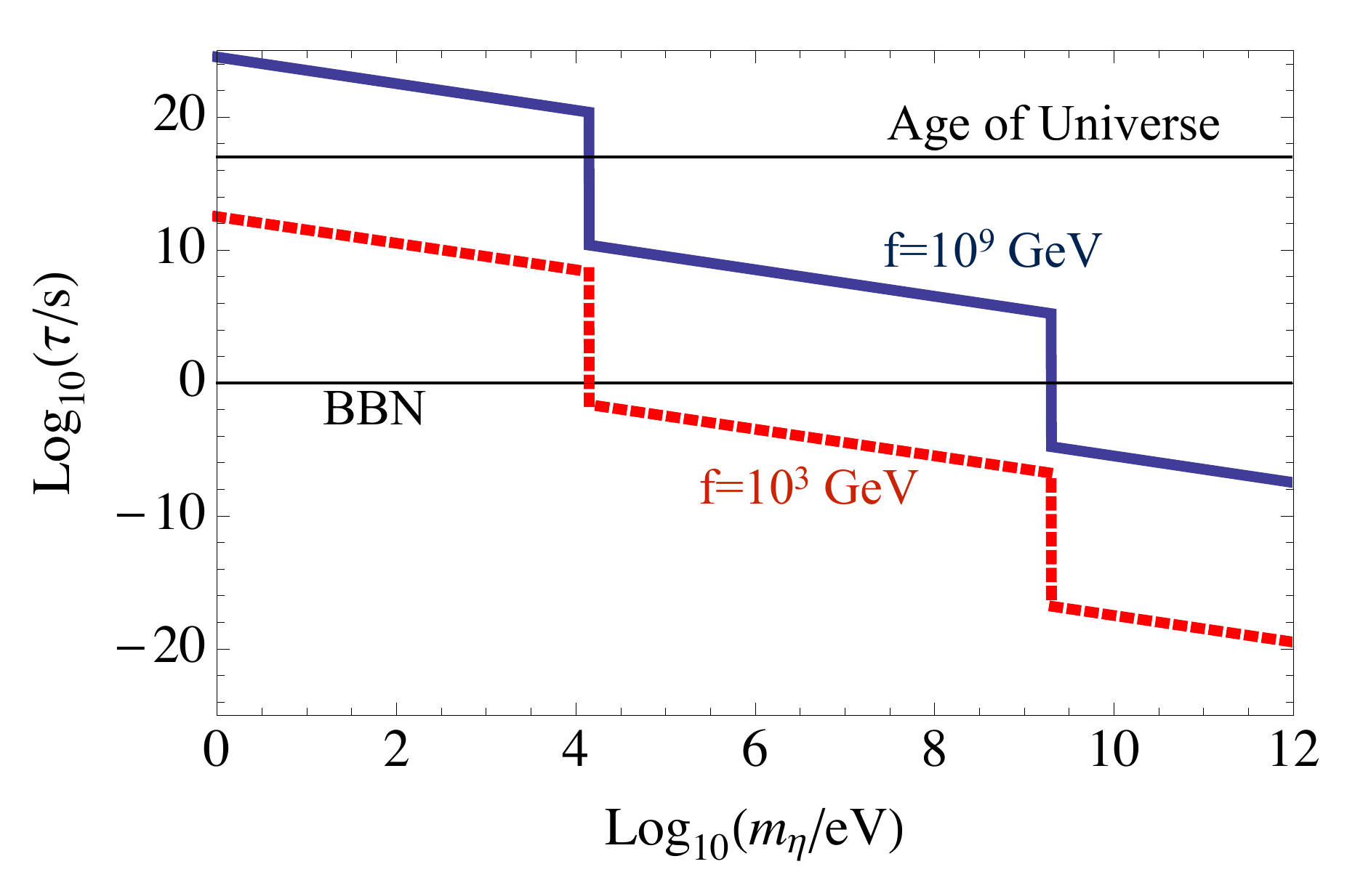}
\caption{\label{fig:rholifetime} Contours of lifetime Log$_{10} (\tau_\eta/s)$ with $M_{N_{2,3}}=1$\,GeV, $M_{N_1} = 7$\,keV for $f=10^9$ GeV (blue solid) and $f=10^3$ GeV (red dotted). The horizontal lines represent the age of the Universe (top) and the time of BBN (bottom).}
\end{figure}

 \textbf{Lifetime:} The $\eta$ lifetime is controlled by decay rates into (both active and sterile) neutrinos. For instance,
\be
\Gamma(\eta\to \nu\nu)\approx\frac{1}{8\pi}\left(\frac{m_\nu}{f}\right)^2\,m_\eta,
\label{eq:rhodecay}
\ee
where $m_\nu\sim 0.1$\,eV is the active neutrino mass scale. For the decay channels $\eta\to N_i\nu$ and $\eta\to N_i N_i$ involving the sterile neutrinos, $m_\nu$ is replaced by $\sqrt{m_{N_i} m_\nu}$ and $m_{N_i}$ respectively. Fig.\,\ref{fig:rholifetime} shows the $\eta$ lifetime as a function of $m_\eta$, with $M_{N_2, N_3}=1$\,GeV and $M_{N_1}=7$\,keV, for two different values of $f$. Depending on the scale $f$ and the available decay channels, a range of interesting lifetimes are possible: $\eta$ can decay before or after BBN (and before/after Cosmic Microwave Background (CMB) decoupling), or live longer than the age of the Universe, providing a potential DM candidate (for studies of majoron DM, see \cite{Rothstein:1992rh,Berezinsky:1993fm,Gu:2010ys,Frigerio:2011in,Queiroz:2014yna,Lattanzi:2014mia,Boucenna:2014uma,Boulebnane:2017fxw,Heeck:2017xbu}).

A pseudo-Goldstone coupling to neutrinos faces several constraints \cite{Choi:1989hi,Pastor:1997nn,Kachelriess:2000qc,Hirsch:2009ee,GarciaiTormo:2011et}. However, many of these constraints weaken/become inapplicable if the pseudo-Goldstone is heavy or can decay into sterile neutrinos. We remark that these constraints are generally not very stringent in the parameter space of interest in our framework.   

\textbf{Cosmology:} In the early Universe, GeV scale sterile neutrinos $N_{2,3}$ (but not the DM candidate $N_1$, which has suppressed couplings to neutrinos) are in equilibrium with the thermal bath due to their mixing with active neutrinos, decouple while relativistic at $T\sim 20$ GeV \cite{Asaka:2006ek}, can grow to dominate the energy density of the Universe, and decay before BBN \cite{Scherrer:1984fd,Bezrukov:2009th,Asaka:2006ek}. 

\begin{figure}[t!]
\includegraphics[width=3.4in, height=1.1in]{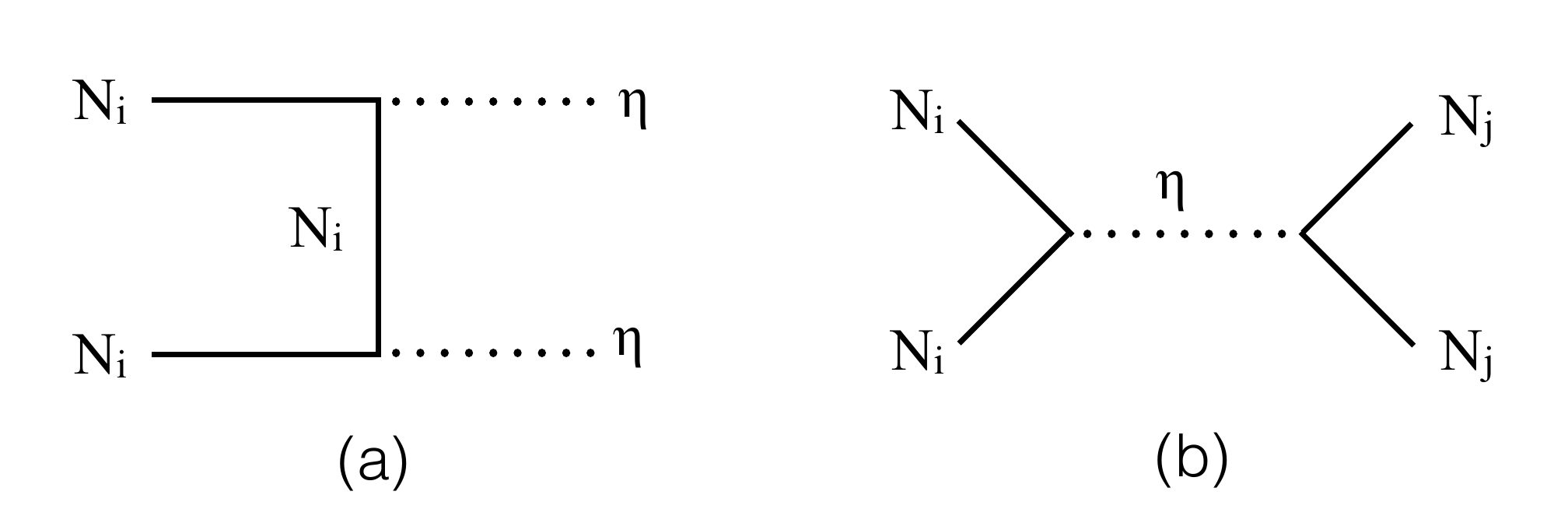}
\caption{\label{fig:diagrams} Sterile neutrino annihilation processes involving the pseudo-Goldstone boson $\eta$.}
\end{figure}

$\eta$ couples appreciably only to the sterile neutrinos, and is produced via sterile neutrino annihilation $N_iN_i\to \eta\eta$ (see Fig.\ref{fig:diagrams}\,(a)) or decay (if kinematically open). The annihilation process, despite $p$-wave suppression, is efficient at high temperatures $T\gtrsim m_{N_{2,3}}$.  The magnitude of $f$ for such annihilations to be rapid can be estimated by comparing the annihilation cross section \cite{Garcia-Cely:2013wda,Gu:2009hn} with the Hubble rate at $T\sim m_{N_{2,3}}$
\be
n_{N_i}\sigma v\sim H ~ \Rightarrow ~ \frac{m_{N_i}^4}{f^4} m_{N_i} \sim \frac{m_{N_i}^2}{M_{Pl}} ~ \Rightarrow ~ f\sim m_{N_i}^{3/4} M_{Pl}^{1/4}.
\label{eq:annihilationcompare}
\ee
For $m_{N_{2,3}}\sim$ GeV, this process is efficient for $f\lesssim 10^5$ GeV, and produces an $\eta$ abundance comparable to the $N_{2,3}$ abundance. For $f>10^5$\,GeV, the annihilation process is feeble, and a small $\eta$ abundance will accumulate via the freeze-in process instead \cite{Chung:1998rq,Hall:2009bx}.

\textbf{Dark Matter Production:} $\eta$ can also mediate $N_iN_i\to N_jN_j$ interactions between the sterile neutrinos (Fig.\ref{fig:diagrams}\,(b)), which enables a novel DM production mechanism $N_iN_i\to N_1N_1$. One can analogously estimate the scale $f$ below which this process \cite{Chacko:2003dt} is efficient:  $f\sim \sqrt{m_{N_1}}(M_{Pl} m_{N_{2,3}})^{1/4}.$ This would generate an $N_1$ abundance comparable to relativistic freezeout, which generally overcloses the Universe, hence this scenario is best avoided. Likewise, $\eta$ decays can also produce DM if $m_\eta\,\textgreater \,2 m_{N_{1}}$. By comparing rates, we find that production from such decays dominates over the annihilation process provided $m_\eta\, \textgreater\, m_{N_{2,3}}^3/f^2$, which generally holds over most of our parameter space. Additional DM production processes, such as $\eta$ annihilation and $N_{2,3}$ decays via an off-shell $\eta$, are always subdominant and therefore neglected. The novel production processes discussed here do not rely on $N_1$ mixing with active neutrinos, which is particularly appealing since this canonical (Dodelson-Widrow) production mechanism \cite{Dodelson:1993je} is now ruled out by various constraints \cite{Boyarsky:2006fg,Boyarsky:2006ag, Boyarsky:2005us,Boyarsky:2007ay,Boyarsky:2007ge,Seljak:2006qw, Asaka:2006nq, Boyarsky:2008xj,Horiuchi:2013noa}. 
 
Next, we discuss various cosmological histories that are possible within this framework. Our purpose is not to provide a comprehensive survey of all possibilities, but simply to highlight some novel and interesting features that can be realized. Since available decay channels and lifetimes are crucial to the cosmological history, we find it useful to organize our discussion into the following three different regimes.

\vskip0.3cm
\noindent\textbf{Heavy regime: $m_\eta\,\textgreater\,m_{N_i}$}
\vskip0.1cm

All $\eta$ decay channels to sterile neutrinos are open, and $\eta$ decays rapidly, long before BBN. If $N_i N_i\to\eta\eta$ is rapid, $\eta$ maintains an equilibrium distribution at $T\gtrsim m_\eta$, and the decay $\eta\to N_1 N_1$ generates a freeze-in abundance of $N_1$, estimated to be \cite{Hall:2009bx,Shakya:2015xnx,Roland:2016gli,Shakya:2016oxf,Merle:2013wta,Adulpravitchai:2014xna,Kang:2014cia,Roland:2014vba,Merle:2015oja}
\be
Y_{eq}\sim 0.1 \frac{M_{Pl}}{m_\eta}\left(\frac{m_{N_1}}{f}\right)^2.
\label{eq:yeq}
\ee 
The observed DM abundance is produced, for instance, with $f\sim 10^5$ GeV, $m_\eta\sim 10$ GeV, and $m_{N_1}\sim 10$ keV.

If the $N_iN_i\to \eta\eta$ annihilation process is feeble, a freeze-in abundance of $\eta$ is generated instead, and its decays produce a small abundance of $N_1$. The $N_1$ yield is suppressed by the branching fraction BR($\eta\to N_1 N_1)=\frac{\Gamma(\eta\to N_1 N_1)}{\Gamma(\eta\to N_{2,3} N_{2,3})}=\frac{m_{N_1}^2}{m_{N_{2,3}}}$. The resulting abundance is much smaller than $Y_{eq}$ from Eq.\,\ref{eq:yeq} and cannot account for all of DM unless $m_{N_1}\sim m_{N_2,N_3}$.


\vskip0.3cm
\noindent\textbf{Intermediate regime: $m_{N_{2,3}}\,\textgreater\,m_\eta\,\textgreater\,m_{N_1}$} 
\vskip0.1cm

\begin{figure}[t!]
\includegraphics[width=3in, height=1.7in]{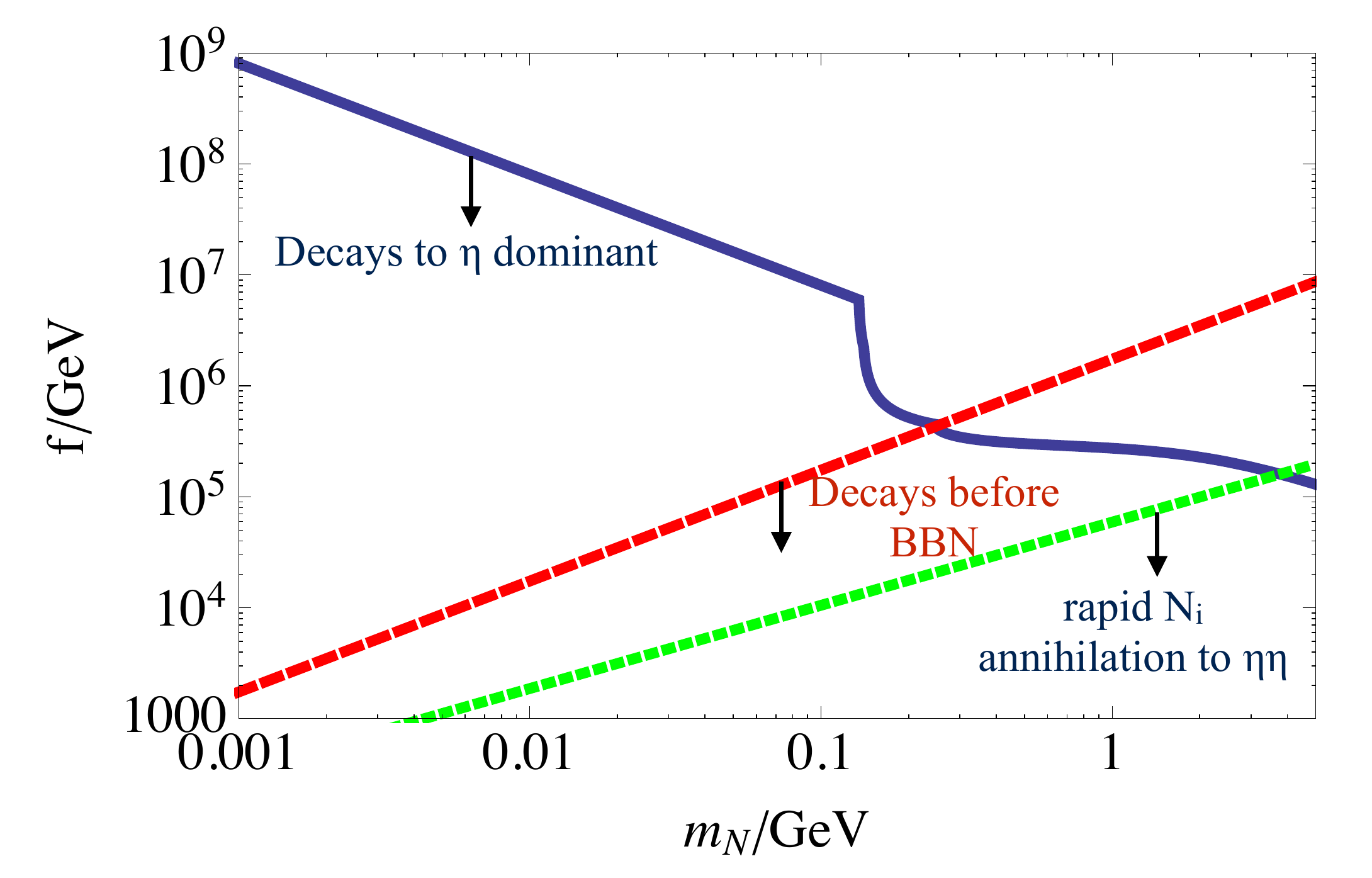}
\caption{\label{fig:widthcompare} Solid blue: Symmetry breaking scale $f$ below which the exotic decay $N\to \eta\nu$ dominates over the standard sterile neutrino decay channels imposed by seesaw relations. Below the dashed red line, this decay channel causes the sterile neutrinos to decay before BBN. Below the dotted green line, sterile neutrino - pseudo-Goldstone interactions are sufficiently rapid to thermalize the two populations in the early Universe. }
\end{figure}

In addition to annihilation processes, $\eta$ can now also be produced directly from heavy sterile neutrino decay. Ignoring phase space suppression, the decay rate is
\be
\Gamma(N_i\to \eta\nu)\approx\frac{1}{16\pi}\frac{m_{N_i} m_\nu}{f^2}\,m_{N_i}\,.
\label{eq:Ndecay}
\ee
If sufficiently large, this exotic decay channel can compete with the standard sterile neutrino decay channels induced by active-sterile mixing \cite{Gorbunov:2007ak}. In Fig.\,\ref{fig:widthcompare}, we plot (blue curve) the scale $f$ below which this channel dominates (assuming standard seesaw relations). In this region, the traditionally searched-for decay modes are suppressed, rendering the sterile neutrinos invisible at detectors such as at DUNE \cite{Adams:2013qkq} and SHiP \cite{Jacobsson:2130433} (unless $N_1$ also decays in the detector, as can occur if it is not DM). 

$N_{2,3}$ are generally required to decay before BBN due to constraints from several recombination era observables \cite{Kusenko:2009up, Hernandez:2014fha,Vincent:2014rja}, necessitating $\tau_{N2,N3}\lesssim 1\,{\rm s}$ and consequently $m_{N2,N3}\gtrsim\mathcal{O}(100)$ MeV in the standard seesaw formalism. The new decay channel $N_i\to \eta\nu$, if dominant, can reduce the sterile neutrino lifetime, allowing lighter masses to be compatible with BBN. In Fig.\,\ref{fig:widthcompare}, the red dashed line shows the scale $f$ below which the sterile neutrino decays before BBN. For $f\lesssim10^{6}$\,GeV, even lighter (MeV scale) sterile neutrinos are compatible with the seesaw as well as BBN constraints, in stark contrast to the standard seesaw requirements. 

Depending on parameters, $\eta$ can decay before or after BBN (Fig.\,\ref{fig:rholifetime}), but its dominant decay channel is to the DM candidate $\eta\to N_1 N_1$. If $N_{2,3}$ decay dominantly into $\eta$, or if $N_1$ thermalizes with $N_{2,3}$, the $N_1$ relic density is overabundant for DM. Viable regions of parameter space instead involve a small fraction of $N_{2,3}$ decaying into $\eta$, which subsequently decays to $N_1$. In this case, $N_1$ accounts for the observed DM abundance (for $m_{N_{2,3}}=1$ GeV) for $f\approx10^9\,\text{GeV} \sqrt{\frac{m_{N_1}}{\text{GeV}}}$. For instance, $m_{N_1}=7$ keV requires $f\sim 10^6$ GeV. 

Here, DM ($N_1$) is produced from late decays of heavier particles ($\eta$ and $N_{2,3}$) and can be warm. Such late production of warm DM can carry interesting cosmological signatures and structure formation implications, which lie beyond the scope of this paper. 

\vskip0.3cm
\noindent\textbf{Light regime: $m_{N_i}\,\textgreater\,m_\eta\,\textgreater\,m_\nu$}
\vskip0.1cm

All sterile neutrinos can now decay into $\eta$. In particular, a new, very long-lived DM decay channel $N_1\to \eta \nu$ emerges. Since $\eta$ subsequently decays into two neutrinos, this can provide distinct signatures at neutrino detectors such as IceCube, Borexino, KamLAND, and Super-Kamiokande. Note that, unlike the standard $N_1\to\gamma\nu$ decay channel, this has no gamma ray counterpart.   

Unlike previous scenarios, $\eta$ is extremely long-lived, and if sufficiently light, can contribute measurably to dark radiation at BBN or CMB \cite{Chikashige:1980qk,Garcia-Cely:2013nin,Garcia-Cely:2013wda}. A Goldstone that freezes out above 100 MeV contributes $\sim0.39$ to $N_{\text{eff}}$ at CMB \cite{Weinberg:2013kea}; this is the case if the sterile neutrino annihilation to $\eta$ is efficient or if sterile neutrinos decay dominantly to $\eta$. If $\eta$ decays after neutrino decoupling, neutrinos from its decays provide additional radiation energy density in the CMB \cite{Chacko:2003dt}.

Finally, if $\eta$ is sufficiently long-lived and heavy, it can also account for part or all of DM. The phenomenology in this case is similar to that of the majoron \cite{Rothstein:1992rh,Berezinsky:1993fm,Gu:2010ys,Queiroz:2014yna, Frigerio:2011in,Lattanzi:2014mia,Boucenna:2014uma,Boulebnane:2017fxw,Heeck:2017xbu}, with neutrino lines as an interesting signal \cite{Garcia-Cely:2017oco}.

\section{Discussion}

We studied the phenomenology of a pseudo-Goldstone boson $\eta$ associated with a spontaneously broken global symmetry in a light (GeV scale) sterile neutrino sector. The presence of sterile neutrinos and $\eta$ at similar mass scales gives rise to several novel possibilities for cosmology, DM, and direct searches. Primary among these are novel sterile neutrino DM production mechanisms from $\eta$-mediated scattering or decay, and new decay channels for heavy sterile neutrinos, which can alleviate BBN bounds and suppress standard search channels at direct search experiments, or provide distinct DM signals at neutrino detectors. Likewise, $\eta$ can contribute measurably to dark radiation at BBN or CMB, inject a late population of SM neutrinos from its late decays, or account for DM. We have only touched upon a few interesting phenomenological possibilities in this framework, and several directions, such as the effect of $\eta$ on leptogenesis \cite{Asaka:2005an,Asaka:2005pn,Asaka:2006nq,JosseMichaux:2011ba,Gu:2009hn}, or differences in the flavor structure and mixing angles from the hidden sector interpretation compared to the canonical seesaw mechanism, could be worthy of further detailed study. 

\medskip
\textit{Acknowledgements: }The authors are supported in part by the DoE under grants DE-SC0007859 and DE-SC0011719.  BS acknowledges support from the University of Cincinnati and thanks the CERN and DESY theory groups, where part of this work was conducted, for hospitality. This work was performed in part at the Aspen Center for Physics, which is supported by National Science Foundation grant PHY-1066293.

\bibliography{exotic_neutrino_bibliography}

\begin{thebibliography}{100}
\expandafter\ifx\csname natexlab\endcsname\relax\def\natexlab#1{#1}\fi
\expandafter\ifx\csname bibnamefont\endcsname\relax
  \def\bibnamefont#1{#1}\fi
\expandafter\ifx\csname bibfnamefont\endcsname\relax
  \def\bibfnamefont#1{#1}\fi
\expandafter\ifx\csname citenamefont\endcsname\relax
  \def\citenamefont#1{#1}\fi
\expandafter\ifx\csname url\endcsname\relax
  \def\url#1{\texttt{#1}}\fi
\expandafter\ifx\csname urlprefix\endcsname\relax\def\urlprefix{URL }\fi
\providecommand{\bibinfo}[2]{#2}
\providecommand{\eprint}[2][]{\url{#2}}

\bibitem[{\citenamefont{Minkowski}(1977)}]{Minkowski:1977sc}
\bibinfo{author}{\bibfnamefont{P.}~\bibnamefont{Minkowski}},
  \bibinfo{journal}{Phys. Lett.} \textbf{\bibinfo{volume}{67B}},
  \bibinfo{pages}{421} (\bibinfo{year}{1977}).

\bibitem[{\citenamefont{Mohapatra and Senjanovic}(1980)}]{Mohapatra:1979ia}
\bibinfo{author}{\bibfnamefont{R.~N.} \bibnamefont{Mohapatra}}
  \bibnamefont{and}
  \bibinfo{author}{\bibfnamefont{G.}~\bibnamefont{Senjanovic}},
  \bibinfo{journal}{Phys. Rev. Lett.} \textbf{\bibinfo{volume}{44}},
  \bibinfo{pages}{912} (\bibinfo{year}{1980}).

\bibitem[{\citenamefont{Mohapatra and Senjanovic}(1981)}]{Mohapatra:1980yp}
\bibinfo{author}{\bibfnamefont{R.~N.} \bibnamefont{Mohapatra}}
  \bibnamefont{and}
  \bibinfo{author}{\bibfnamefont{G.}~\bibnamefont{Senjanovic}},
  \bibinfo{journal}{Phys. Rev.} \textbf{\bibinfo{volume}{D23}},
  \bibinfo{pages}{165} (\bibinfo{year}{1981}).

\bibitem[{\citenamefont{Yanagida}(1980)}]{Yanagida:1980xy}
\bibinfo{author}{\bibfnamefont{T.}~\bibnamefont{Yanagida}},
  \bibinfo{journal}{Prog. Theor. Phys.} \textbf{\bibinfo{volume}{64}},
  \bibinfo{pages}{1103} (\bibinfo{year}{1980}).

\bibitem[{\citenamefont{Gell-Mann et~al.}(1979)\citenamefont{Gell-Mann, Ramond,
  and Slansky}}]{GellMann:1980vs}
\bibinfo{author}{\bibfnamefont{M.}~\bibnamefont{Gell-Mann}},
  \bibinfo{author}{\bibfnamefont{P.}~\bibnamefont{Ramond}}, \bibnamefont{and}
  \bibinfo{author}{\bibfnamefont{R.}~\bibnamefont{Slansky}},
  \bibinfo{journal}{Conf. Proc.} \textbf{\bibinfo{volume}{C790927}},
  \bibinfo{pages}{315} (\bibinfo{year}{1979}), \eprint{1306.4669}.

\bibitem[{\citenamefont{Schechter and Valle}(1980)}]{Schechter:1980gr}
\bibinfo{author}{\bibfnamefont{J.}~\bibnamefont{Schechter}} \bibnamefont{and}
  \bibinfo{author}{\bibfnamefont{J.~W.~F.} \bibnamefont{Valle}},
  \bibinfo{journal}{Phys. Rev.} \textbf{\bibinfo{volume}{D22}},
  \bibinfo{pages}{2227} (\bibinfo{year}{1980}).

\bibitem[{\citenamefont{Asaka et~al.}(2005)\citenamefont{Asaka, Blanchet, and
  Shaposhnikov}}]{Asaka:2005an}
\bibinfo{author}{\bibfnamefont{T.}~\bibnamefont{Asaka}},
  \bibinfo{author}{\bibfnamefont{S.}~\bibnamefont{Blanchet}}, \bibnamefont{and}
  \bibinfo{author}{\bibfnamefont{M.}~\bibnamefont{Shaposhnikov}},
  \bibinfo{journal}{Phys. Lett.} \textbf{\bibinfo{volume}{B631}},
  \bibinfo{pages}{151} (\bibinfo{year}{2005}), \eprint{hep-ph/0503065}.

\bibitem[{\citenamefont{Asaka and Shaposhnikov}(2005)}]{Asaka:2005pn}
\bibinfo{author}{\bibfnamefont{T.}~\bibnamefont{Asaka}} \bibnamefont{and}
  \bibinfo{author}{\bibfnamefont{M.}~\bibnamefont{Shaposhnikov}},
  \bibinfo{journal}{Phys. Lett.} \textbf{\bibinfo{volume}{B620}},
  \bibinfo{pages}{17} (\bibinfo{year}{2005}), \eprint{hep-ph/0505013}.

\bibitem[{\citenamefont{Asaka et~al.}(2007)\citenamefont{Asaka, Laine, and
  Shaposhnikov}}]{Asaka:2006nq}
\bibinfo{author}{\bibfnamefont{T.}~\bibnamefont{Asaka}},
  \bibinfo{author}{\bibfnamefont{M.}~\bibnamefont{Laine}}, \bibnamefont{and}
  \bibinfo{author}{\bibfnamefont{M.}~\bibnamefont{Shaposhnikov}},
  \bibinfo{journal}{JHEP} \textbf{\bibinfo{volume}{01}}, \bibinfo{pages}{091}
  (\bibinfo{year}{2007}), \bibinfo{note}{[Erratum: JHEP02,028(2015)]},
  \eprint{hep-ph/0612182}.

\bibitem[{\citenamefont{Abazajian et~al.}(2012)}]{Abazajian:2012ys}
\bibinfo{author}{\bibfnamefont{K.~N.} \bibnamefont{Abazajian}}
  \bibnamefont{et~al.} (\bibinfo{year}{2012}), \eprint{1204.5379}.

\bibitem[{\citenamefont{Drewes et~al.}(2017)}]{Adhikari:2016bei}
\bibinfo{author}{\bibfnamefont{M.}~\bibnamefont{Drewes}} \bibnamefont{et~al.},
  \bibinfo{journal}{JCAP} \textbf{\bibinfo{volume}{1701}}, \bibinfo{pages}{025}
  (\bibinfo{year}{2017}), \eprint{1602.04816}.

\bibitem[{\citenamefont{Ma}(2017)}]{Ma:2017dbl}
\bibinfo{author}{\bibfnamefont{E.}~\bibnamefont{Ma}}, \bibinfo{journal}{Mod.
  Phys. Lett.} \textbf{\bibinfo{volume}{A32}}, \bibinfo{pages}{1730007}
  (\bibinfo{year}{2017}), \eprint{1702.03281}.

\bibitem[{\citenamefont{Alanne et~al.}(2017)\citenamefont{Alanne, Meroni, and
  Tuominen}}]{Alanne:2017sip}
\bibinfo{author}{\bibfnamefont{T.}~\bibnamefont{Alanne}},
  \bibinfo{author}{\bibfnamefont{A.}~\bibnamefont{Meroni}}, \bibnamefont{and}
  \bibinfo{author}{\bibfnamefont{K.}~\bibnamefont{Tuominen}},
  \bibinfo{journal}{Phys. Rev.} \textbf{\bibinfo{volume}{D96}},
  \bibinfo{pages}{095015} (\bibinfo{year}{2017}), \eprint{1706.10128}.

\bibitem[{\citenamefont{Sayre et~al.}(2005)\citenamefont{Sayre, Wiesenfeldt,
  and Willenbrock}}]{Sayre:2005yh}
\bibinfo{author}{\bibfnamefont{J.}~\bibnamefont{Sayre}},
  \bibinfo{author}{\bibfnamefont{S.}~\bibnamefont{Wiesenfeldt}},
  \bibnamefont{and}
  \bibinfo{author}{\bibfnamefont{S.}~\bibnamefont{Willenbrock}},
  \bibinfo{journal}{Phys. Rev.} \textbf{\bibinfo{volume}{D72}},
  \bibinfo{pages}{015001} (\bibinfo{year}{2005}), \eprint{hep-ph/0504198}.

\bibitem[{\citenamefont{Chikashige et~al.}(1981)\citenamefont{Chikashige,
  Mohapatra, and Peccei}}]{Chikashige:1980ui}
\bibinfo{author}{\bibfnamefont{Y.}~\bibnamefont{Chikashige}},
  \bibinfo{author}{\bibfnamefont{R.~N.} \bibnamefont{Mohapatra}},
  \bibnamefont{and} \bibinfo{author}{\bibfnamefont{R.~D.}
  \bibnamefont{Peccei}}, \bibinfo{journal}{Phys. Lett.}
  \textbf{\bibinfo{volume}{98B}}, \bibinfo{pages}{265} (\bibinfo{year}{1981}).

\bibitem[{\citenamefont{Gelmini and Roncadelli}(1981)}]{Gelmini:1980re}
\bibinfo{author}{\bibfnamefont{G.~B.} \bibnamefont{Gelmini}} \bibnamefont{and}
  \bibinfo{author}{\bibfnamefont{M.}~\bibnamefont{Roncadelli}},
  \bibinfo{journal}{Phys. Lett.} \textbf{\bibinfo{volume}{99B}},
  \bibinfo{pages}{411} (\bibinfo{year}{1981}).

\bibitem[{\citenamefont{Georgi et~al.}(1981)\citenamefont{Georgi, Glashow, and
  Nussinov}}]{Georgi:1981pg}
\bibinfo{author}{\bibfnamefont{H.~M.} \bibnamefont{Georgi}},
  \bibinfo{author}{\bibfnamefont{S.~L.} \bibnamefont{Glashow}},
  \bibnamefont{and} \bibinfo{author}{\bibfnamefont{S.}~\bibnamefont{Nussinov}},
  \bibinfo{journal}{Nucl. Phys.} \textbf{\bibinfo{volume}{B193}},
  \bibinfo{pages}{297} (\bibinfo{year}{1981}).

\bibitem[{\citenamefont{Schechter and Valle}(1982)}]{Schechter:1981cv}
\bibinfo{author}{\bibfnamefont{J.}~\bibnamefont{Schechter}} \bibnamefont{and}
  \bibinfo{author}{\bibfnamefont{J.~W.~F.} \bibnamefont{Valle}},
  \bibinfo{journal}{Phys. Rev.} \textbf{\bibinfo{volume}{D25}},
  \bibinfo{pages}{774} (\bibinfo{year}{1982}).

\bibitem[{\citenamefont{Gelmini et~al.}(1982)\citenamefont{Gelmini, Nussinov,
  and Roncadelli}}]{Gelmini:1982rr}
\bibinfo{author}{\bibfnamefont{G.~B.} \bibnamefont{Gelmini}},
  \bibinfo{author}{\bibfnamefont{S.}~\bibnamefont{Nussinov}}, \bibnamefont{and}
  \bibinfo{author}{\bibfnamefont{M.}~\bibnamefont{Roncadelli}},
  \bibinfo{journal}{Nucl. Phys.} \textbf{\bibinfo{volume}{B209}},
  \bibinfo{pages}{157} (\bibinfo{year}{1982}).

\bibitem[{\citenamefont{Lindner et~al.}(2011)\citenamefont{Lindner, Schmidt,
  and Schwetz}}]{Lindner:2011it}
\bibinfo{author}{\bibfnamefont{M.}~\bibnamefont{Lindner}},
  \bibinfo{author}{\bibfnamefont{D.}~\bibnamefont{Schmidt}}, \bibnamefont{and}
  \bibinfo{author}{\bibfnamefont{T.}~\bibnamefont{Schwetz}},
  \bibinfo{journal}{Phys. Lett.} \textbf{\bibinfo{volume}{B705}},
  \bibinfo{pages}{324} (\bibinfo{year}{2011}), \eprint{1105.4626}.

\bibitem[{\citenamefont{Escudero et~al.}(2017)\citenamefont{Escudero, Rius, and
  Sanz}}]{Escudero:2016tzx}
\bibinfo{author}{\bibfnamefont{M.}~\bibnamefont{Escudero}},
  \bibinfo{author}{\bibfnamefont{N.}~\bibnamefont{Rius}}, \bibnamefont{and}
  \bibinfo{author}{\bibfnamefont{V.}~\bibnamefont{Sanz}},
  \bibinfo{journal}{JHEP} \textbf{\bibinfo{volume}{02}}, \bibinfo{pages}{045}
  (\bibinfo{year}{2017}), \eprint{1606.01258}.

\bibitem[{\citenamefont{Maiezza et~al.}(2015)\citenamefont{Maiezza, Nemevšek,
  and Nesti}}]{Maiezza:2015lza}
\bibinfo{author}{\bibfnamefont{A.}~\bibnamefont{Maiezza}},
  \bibinfo{author}{\bibfnamefont{M.}~\bibnamefont{Nemevšek}},
  \bibnamefont{and} \bibinfo{author}{\bibfnamefont{F.}~\bibnamefont{Nesti}},
  \bibinfo{journal}{Phys. Rev. Lett.} \textbf{\bibinfo{volume}{115}},
  \bibinfo{pages}{081802} (\bibinfo{year}{2015}), \eprint{1503.06834}.

\bibitem[{\citenamefont{Nemevšek et~al.}(2017)\citenamefont{Nemevšek, Nesti,
  and Vasquez}}]{Nemevsek:2016enw}
\bibinfo{author}{\bibfnamefont{M.}~\bibnamefont{Nemevšek}},
  \bibinfo{author}{\bibfnamefont{F.}~\bibnamefont{Nesti}}, \bibnamefont{and}
  \bibinfo{author}{\bibfnamefont{J.~C.} \bibnamefont{Vasquez}},
  \bibinfo{journal}{JHEP} \textbf{\bibinfo{volume}{04}}, \bibinfo{pages}{114}
  (\bibinfo{year}{2017}), \eprint{1612.06840}.

\bibitem[{\citenamefont{Maiezza et~al.}(2017)\citenamefont{Maiezza,
  Senjanović, and Vasquez}}]{Maiezza:2016ybz}
\bibinfo{author}{\bibfnamefont{A.}~\bibnamefont{Maiezza}},
  \bibinfo{author}{\bibfnamefont{G.}~\bibnamefont{Senjanović}},
  \bibnamefont{and} \bibinfo{author}{\bibfnamefont{J.~C.}
  \bibnamefont{Vasquez}}, \bibinfo{journal}{Phys. Rev.}
  \textbf{\bibinfo{volume}{D95}}, \bibinfo{pages}{095004}
  (\bibinfo{year}{2017}), \eprint{1612.09146}.

\bibitem[{\citenamefont{Dev et~al.}(2017)\citenamefont{Dev, Mohapatra, and
  Zhang}}]{Dev:2017dui}
\bibinfo{author}{\bibfnamefont{P.~S.~B.} \bibnamefont{Dev}},
  \bibinfo{author}{\bibfnamefont{R.~N.} \bibnamefont{Mohapatra}},
  \bibnamefont{and} \bibinfo{author}{\bibfnamefont{Y.}~\bibnamefont{Zhang}},
  \bibinfo{journal}{Nucl. Phys.} \textbf{\bibinfo{volume}{B923}},
  \bibinfo{pages}{179} (\bibinfo{year}{2017}), \eprint{1703.02471}.

\bibitem[{\citenamefont{Keung and Senjanovic}(1983)}]{Keung:1983uu}
\bibinfo{author}{\bibfnamefont{W.-Y.} \bibnamefont{Keung}} \bibnamefont{and}
  \bibinfo{author}{\bibfnamefont{G.}~\bibnamefont{Senjanovic}},
  \bibinfo{journal}{Phys. Rev. Lett.} \textbf{\bibinfo{volume}{50}},
  \bibinfo{pages}{1427} (\bibinfo{year}{1983}).

\bibitem[{\citenamefont{Asaka et~al.}(2006)\citenamefont{Asaka, Shaposhnikov,
  and Kusenko}}]{Asaka:2006ek}
\bibinfo{author}{\bibfnamefont{T.}~\bibnamefont{Asaka}},
  \bibinfo{author}{\bibfnamefont{M.}~\bibnamefont{Shaposhnikov}},
  \bibnamefont{and} \bibinfo{author}{\bibfnamefont{A.}~\bibnamefont{Kusenko}},
  \bibinfo{journal}{Phys. Lett.} \textbf{\bibinfo{volume}{B638}},
  \bibinfo{pages}{401} (\bibinfo{year}{2006}), \eprint{hep-ph/0602150}.

\bibitem[{\citenamefont{Cleaver et~al.}(1998)\citenamefont{Cleaver, Cvetic,
  Espinosa, Everett, and Langacker}}]{Cleaver:1997nj}
\bibinfo{author}{\bibfnamefont{G.}~\bibnamefont{Cleaver}},
  \bibinfo{author}{\bibfnamefont{M.}~\bibnamefont{Cvetic}},
  \bibinfo{author}{\bibfnamefont{J.~R.} \bibnamefont{Espinosa}},
  \bibinfo{author}{\bibfnamefont{L.~L.} \bibnamefont{Everett}},
  \bibnamefont{and}
  \bibinfo{author}{\bibfnamefont{P.}~\bibnamefont{Langacker}},
  \bibinfo{journal}{Phys. Rev.} \textbf{\bibinfo{volume}{D57}},
  \bibinfo{pages}{2701} (\bibinfo{year}{1998}), \eprint{hep-ph/9705391}.

\bibitem[{\citenamefont{Langacker}(1998)}]{Langacker:1998ut}
\bibinfo{author}{\bibfnamefont{P.}~\bibnamefont{Langacker}},
  \bibinfo{journal}{Phys. Rev.} \textbf{\bibinfo{volume}{D58}},
  \bibinfo{pages}{093017} (\bibinfo{year}{1998}), \eprint{hep-ph/9805281}.

\bibitem[{\citenamefont{Arkani-Hamed et~al.}(2001)\citenamefont{Arkani-Hamed,
  Hall, Murayama, Tucker-Smith, and Weiner}}]{ArkaniHamed:2000bq}
\bibinfo{author}{\bibfnamefont{N.}~\bibnamefont{Arkani-Hamed}},
  \bibinfo{author}{\bibfnamefont{L.~J.} \bibnamefont{Hall}},
  \bibinfo{author}{\bibfnamefont{H.}~\bibnamefont{Murayama}},
  \bibinfo{author}{\bibfnamefont{D.}~\bibnamefont{Tucker-Smith}},
  \bibnamefont{and} \bibinfo{author}{\bibfnamefont{N.}~\bibnamefont{Weiner}},
  \bibinfo{journal}{Phys. Rev.} \textbf{\bibinfo{volume}{D64}},
  \bibinfo{pages}{115011} (\bibinfo{year}{2001}), \eprint{hep-ph/0006312}.

\bibitem[{\citenamefont{Arkani-Hamed et~al.}(2000)\citenamefont{Arkani-Hamed,
  Hall, Murayama, Tucker-Smith, and Weiner}}]{ArkaniHamed:2000kj}
\bibinfo{author}{\bibfnamefont{N.}~\bibnamefont{Arkani-Hamed}},
  \bibinfo{author}{\bibfnamefont{L.~J.} \bibnamefont{Hall}},
  \bibinfo{author}{\bibfnamefont{H.}~\bibnamefont{Murayama}},
  \bibinfo{author}{\bibfnamefont{D.}~\bibnamefont{Tucker-Smith}},
  \bibnamefont{and} \bibinfo{author}{\bibfnamefont{N.}~\bibnamefont{Weiner}}
  (\bibinfo{year}{2000}), \eprint{hep-ph/0007001}.

\bibitem[{\citenamefont{Wells}(2005)}]{Wells:2004di}
\bibinfo{author}{\bibfnamefont{J.~D.} \bibnamefont{Wells}},
  \bibinfo{journal}{Phys. Rev.} \textbf{\bibinfo{volume}{D71}},
  \bibinfo{pages}{015013} (\bibinfo{year}{2005}), \eprint{hep-ph/0411041}.

\bibitem[{\citenamefont{Roland et~al.}(2015{\natexlab{a}})\citenamefont{Roland,
  Shakya, and Wells}}]{Roland:2014vba}
\bibinfo{author}{\bibfnamefont{S.~B.} \bibnamefont{Roland}},
  \bibinfo{author}{\bibfnamefont{B.}~\bibnamefont{Shakya}}, \bibnamefont{and}
  \bibinfo{author}{\bibfnamefont{J.~D.} \bibnamefont{Wells}},
  \bibinfo{journal}{Phys. Rev.} \textbf{\bibinfo{volume}{D92}},
  \bibinfo{pages}{113009} (\bibinfo{year}{2015}{\natexlab{a}}),
  \eprint{1412.4791}.

\bibitem[{\citenamefont{Roland et~al.}(2015{\natexlab{b}})\citenamefont{Roland,
  Shakya, and Wells}}]{Roland:2015yoa}
\bibinfo{author}{\bibfnamefont{S.~B.} \bibnamefont{Roland}},
  \bibinfo{author}{\bibfnamefont{B.}~\bibnamefont{Shakya}}, \bibnamefont{and}
  \bibinfo{author}{\bibfnamefont{J.~D.} \bibnamefont{Wells}},
  \bibinfo{journal}{Phys. Rev.} \textbf{\bibinfo{volume}{D92}},
  \bibinfo{pages}{095018} (\bibinfo{year}{2015}{\natexlab{b}}),
  \eprint{1506.08195}.

\bibitem[{\citenamefont{Roland and Shakya}(2016)}]{Roland:2016gli}
\bibinfo{author}{\bibfnamefont{S.~B.} \bibnamefont{Roland}} \bibnamefont{and}
  \bibinfo{author}{\bibfnamefont{B.}~\bibnamefont{Shakya}}
  (\bibinfo{year}{2016}), \eprint{1609.06739}.

\bibitem[{\citenamefont{Falkowski et~al.}(2009)\citenamefont{Falkowski,
  Juknevich, and Shelton}}]{Falkowski:2009yz}
\bibinfo{author}{\bibfnamefont{A.}~\bibnamefont{Falkowski}},
  \bibinfo{author}{\bibfnamefont{J.}~\bibnamefont{Juknevich}},
  \bibnamefont{and} \bibinfo{author}{\bibfnamefont{J.}~\bibnamefont{Shelton}}
  (\bibinfo{year}{2009}), \eprint{0908.1790}.

\bibitem[{\citenamefont{Falkowski et~al.}(2011)\citenamefont{Falkowski,
  Ruderman, and Volansky}}]{Falkowski:2011xh}
\bibinfo{author}{\bibfnamefont{A.}~\bibnamefont{Falkowski}},
  \bibinfo{author}{\bibfnamefont{J.~T.} \bibnamefont{Ruderman}},
  \bibnamefont{and} \bibinfo{author}{\bibfnamefont{T.}~\bibnamefont{Volansky}},
  \bibinfo{journal}{JHEP} \textbf{\bibinfo{volume}{05}}, \bibinfo{pages}{106}
  (\bibinfo{year}{2011}), \eprint{1101.4936}.

\bibitem[{\citenamefont{Pospelov}(2011)}]{Pospelov:2011ha}
\bibinfo{author}{\bibfnamefont{M.}~\bibnamefont{Pospelov}},
  \bibinfo{journal}{Phys. Rev.} \textbf{\bibinfo{volume}{D84}},
  \bibinfo{pages}{085008} (\bibinfo{year}{2011}), \eprint{1103.3261}.

\bibitem[{\citenamefont{Pospelov and Pradler}(2012)}]{Pospelov:2012gm}
\bibinfo{author}{\bibfnamefont{M.}~\bibnamefont{Pospelov}} \bibnamefont{and}
  \bibinfo{author}{\bibfnamefont{J.}~\bibnamefont{Pradler}},
  \bibinfo{journal}{Phys. Rev.} \textbf{\bibinfo{volume}{D85}},
  \bibinfo{pages}{113016} (\bibinfo{year}{2012}), \bibinfo{note}{[Erratum:
  Phys. Rev.D88,no.3,039904(2013)]}, \eprint{1203.0545}.

\bibitem[{\citenamefont{Cherry et~al.}(2014)\citenamefont{Cherry, Friedland,
  and Shoemaker}}]{Cherry:2014xra}
\bibinfo{author}{\bibfnamefont{J.~F.} \bibnamefont{Cherry}},
  \bibinfo{author}{\bibfnamefont{A.}~\bibnamefont{Friedland}},
  \bibnamefont{and} \bibinfo{author}{\bibfnamefont{I.~M.}
  \bibnamefont{Shoemaker}} (\bibinfo{year}{2014}), \eprint{1411.1071}.

\bibitem[{\citenamefont{Berryman et~al.}(2017)\citenamefont{Berryman,
  de~Gouvêa, Kelly, and Zhang}}]{Berryman:2017twh}
\bibinfo{author}{\bibfnamefont{J.~M.} \bibnamefont{Berryman}},
  \bibinfo{author}{\bibfnamefont{A.}~\bibnamefont{de~Gouvêa}},
  \bibinfo{author}{\bibfnamefont{K.~J.} \bibnamefont{Kelly}}, \bibnamefont{and}
  \bibinfo{author}{\bibfnamefont{Y.}~\bibnamefont{Zhang}},
  \bibinfo{journal}{Phys. Rev.} \textbf{\bibinfo{volume}{D96}},
  \bibinfo{pages}{075010} (\bibinfo{year}{2017}), \eprint{1706.02722}.

\bibitem[{\citenamefont{Batell et~al.}(2017)\citenamefont{Batell, Han, McKeen,
  and Shams Es~Haghi}}]{Batell:2017cmf}
\bibinfo{author}{\bibfnamefont{B.}~\bibnamefont{Batell}},
  \bibinfo{author}{\bibfnamefont{T.}~\bibnamefont{Han}},
  \bibinfo{author}{\bibfnamefont{D.}~\bibnamefont{McKeen}}, \bibnamefont{and}
  \bibinfo{author}{\bibfnamefont{B.}~\bibnamefont{Shams Es~Haghi}}
  (\bibinfo{year}{2017}), \eprint{1709.07001}.

\bibitem[{\citenamefont{Schmaltz and Weiner}(2017)}]{Schmaltz:2017oov}
\bibinfo{author}{\bibfnamefont{M.}~\bibnamefont{Schmaltz}} \bibnamefont{and}
  \bibinfo{author}{\bibfnamefont{N.}~\bibnamefont{Weiner}}
  (\bibinfo{year}{2017}), \eprint{1709.09164}.

\bibitem[{\citenamefont{Chun et~al.}(1995)\citenamefont{Chun, Joshipura, and
  Smirnov}}]{Chun:1995js}
\bibinfo{author}{\bibfnamefont{E.~J.} \bibnamefont{Chun}},
  \bibinfo{author}{\bibfnamefont{A.~S.} \bibnamefont{Joshipura}},
  \bibnamefont{and} \bibinfo{author}{\bibfnamefont{A.~{\relax Yu}.}
  \bibnamefont{Smirnov}}, \bibinfo{journal}{Phys. Lett.}
  \textbf{\bibinfo{volume}{B357}}, \bibinfo{pages}{608} (\bibinfo{year}{1995}),
  \eprint{hep-ph/9505275}.

\bibitem[{\citenamefont{Ma and Roy}(1995)}]{Ma:1995gf}
\bibinfo{author}{\bibfnamefont{E.}~\bibnamefont{Ma}} \bibnamefont{and}
  \bibinfo{author}{\bibfnamefont{P.}~\bibnamefont{Roy}},
  \bibinfo{journal}{Phys. Rev.} \textbf{\bibinfo{volume}{D52}},
  \bibinfo{pages}{R4780} (\bibinfo{year}{1995}), \eprint{hep-ph/9504342}.

\bibitem[{\citenamefont{Zhang}(2012)}]{Zhang:2011vh}
\bibinfo{author}{\bibfnamefont{H.}~\bibnamefont{Zhang}},
  \bibinfo{journal}{Phys. Lett.} \textbf{\bibinfo{volume}{B714}},
  \bibinfo{pages}{262} (\bibinfo{year}{2012}), \eprint{1110.6838}.

\bibitem[{\citenamefont{Barry et~al.}(2011)\citenamefont{Barry, Rodejohann, and
  Zhang}}]{Barry:2011wb}
\bibinfo{author}{\bibfnamefont{J.}~\bibnamefont{Barry}},
  \bibinfo{author}{\bibfnamefont{W.}~\bibnamefont{Rodejohann}},
  \bibnamefont{and} \bibinfo{author}{\bibfnamefont{H.}~\bibnamefont{Zhang}},
  \bibinfo{journal}{JHEP} \textbf{\bibinfo{volume}{07}}, \bibinfo{pages}{091}
  (\bibinfo{year}{2011}), \eprint{1105.3911}.

\bibitem[{\citenamefont{Boulebnane et~al.}(2017)\citenamefont{Boulebnane,
  Heeck, Nguyen, and Teresi}}]{Boulebnane:2017fxw}
\bibinfo{author}{\bibfnamefont{S.}~\bibnamefont{Boulebnane}},
  \bibinfo{author}{\bibfnamefont{J.}~\bibnamefont{Heeck}},
  \bibinfo{author}{\bibfnamefont{A.}~\bibnamefont{Nguyen}}, \bibnamefont{and}
  \bibinfo{author}{\bibfnamefont{D.}~\bibnamefont{Teresi}}
  (\bibinfo{year}{2017}), \eprint{1709.07283}.

\bibitem[{\citenamefont{Rothstein et~al.}(1993)\citenamefont{Rothstein, Babu,
  and Seckel}}]{Rothstein:1992rh}
\bibinfo{author}{\bibfnamefont{I.~Z.} \bibnamefont{Rothstein}},
  \bibinfo{author}{\bibfnamefont{K.~S.} \bibnamefont{Babu}}, \bibnamefont{and}
  \bibinfo{author}{\bibfnamefont{D.}~\bibnamefont{Seckel}},
  \bibinfo{journal}{Nucl. Phys.} \textbf{\bibinfo{volume}{B403}},
  \bibinfo{pages}{725} (\bibinfo{year}{1993}), \eprint{hep-ph/9301213}.

\bibitem[{\citenamefont{Akhmedov et~al.}(1993)\citenamefont{Akhmedov,
  Berezhiani, Mohapatra, and Senjanovic}}]{Akhmedov:1992hi}
\bibinfo{author}{\bibfnamefont{E.~K.} \bibnamefont{Akhmedov}},
  \bibinfo{author}{\bibfnamefont{Z.~G.} \bibnamefont{Berezhiani}},
  \bibinfo{author}{\bibfnamefont{R.~N.} \bibnamefont{Mohapatra}},
  \bibnamefont{and}
  \bibinfo{author}{\bibfnamefont{G.}~\bibnamefont{Senjanovic}},
  \bibinfo{journal}{Phys. Lett.} \textbf{\bibinfo{volume}{B299}},
  \bibinfo{pages}{90} (\bibinfo{year}{1993}), \eprint{hep-ph/9209285}.

\bibitem[{\citenamefont{Frigerio et~al.}(2011)\citenamefont{Frigerio, Hambye,
  and Masso}}]{Frigerio:2011in}
\bibinfo{author}{\bibfnamefont{M.}~\bibnamefont{Frigerio}},
  \bibinfo{author}{\bibfnamefont{T.}~\bibnamefont{Hambye}}, \bibnamefont{and}
  \bibinfo{author}{\bibfnamefont{E.}~\bibnamefont{Masso}},
  \bibinfo{journal}{Phys. Rev.} \textbf{\bibinfo{volume}{X1}},
  \bibinfo{pages}{021026} (\bibinfo{year}{2011}), \eprint{1107.4564}.

\bibitem[{\citenamefont{Gelmini et~al.}(1983)\citenamefont{Gelmini, Nussinov,
  and Yanagida}}]{Gelmini:1982zz}
\bibinfo{author}{\bibfnamefont{G.~B.} \bibnamefont{Gelmini}},
  \bibinfo{author}{\bibfnamefont{S.}~\bibnamefont{Nussinov}}, \bibnamefont{and}
  \bibinfo{author}{\bibfnamefont{T.}~\bibnamefont{Yanagida}},
  \bibinfo{journal}{Nucl. Phys.} \textbf{\bibinfo{volume}{B219}},
  \bibinfo{pages}{31} (\bibinfo{year}{1983}).

\bibitem[{\citenamefont{Jungman and Luty}(1991)}]{Jungman:1991sh}
\bibinfo{author}{\bibfnamefont{G.}~\bibnamefont{Jungman}} \bibnamefont{and}
  \bibinfo{author}{\bibfnamefont{M.~A.} \bibnamefont{Luty}},
  \bibinfo{journal}{Nucl. Phys.} \textbf{\bibinfo{volume}{B361}},
  \bibinfo{pages}{24} (\bibinfo{year}{1991}).

\bibitem[{\citenamefont{Pilaftsis}(1994)}]{Pilaftsis:1993af}
\bibinfo{author}{\bibfnamefont{A.}~\bibnamefont{Pilaftsis}},
  \bibinfo{journal}{Phys. Rev.} \textbf{\bibinfo{volume}{D49}},
  \bibinfo{pages}{2398} (\bibinfo{year}{1994}), \eprint{hep-ph/9308258}.

\bibitem[{\citenamefont{Garcia-Cely and Heeck}(2017)}]{Garcia-Cely:2017oco}
\bibinfo{author}{\bibfnamefont{C.}~\bibnamefont{Garcia-Cely}} \bibnamefont{and}
  \bibinfo{author}{\bibfnamefont{J.}~\bibnamefont{Heeck}},
  \bibinfo{journal}{JHEP} \textbf{\bibinfo{volume}{05}}, \bibinfo{pages}{102}
  (\bibinfo{year}{2017}), \eprint{1701.07209}.

\bibitem[{\citenamefont{Bulbul et~al.}(2014)\citenamefont{Bulbul, Markevitch,
  Foster, Smith, Loewenstein et~al.}}]{Bulbul:2014sua}
\bibinfo{author}{\bibfnamefont{E.}~\bibnamefont{Bulbul}},
  \bibinfo{author}{\bibfnamefont{M.}~\bibnamefont{Markevitch}},
  \bibinfo{author}{\bibfnamefont{A.}~\bibnamefont{Foster}},
  \bibinfo{author}{\bibfnamefont{R.~K.} \bibnamefont{Smith}},
  \bibinfo{author}{\bibfnamefont{M.}~\bibnamefont{Loewenstein}},
  \bibnamefont{et~al.}, \bibinfo{journal}{Astrophys.J.}
  \textbf{\bibinfo{volume}{789}}, \bibinfo{pages}{13} (\bibinfo{year}{2014}),
  \eprint{1402.2301}.

\bibitem[{\citenamefont{Boyarsky et~al.}(2014)\citenamefont{Boyarsky,
  Ruchayskiy, Iakubovskyi, and Franse}}]{Boyarsky:2014jta}
\bibinfo{author}{\bibfnamefont{A.}~\bibnamefont{Boyarsky}},
  \bibinfo{author}{\bibfnamefont{O.}~\bibnamefont{Ruchayskiy}},
  \bibinfo{author}{\bibfnamefont{D.}~\bibnamefont{Iakubovskyi}},
  \bibnamefont{and} \bibinfo{author}{\bibfnamefont{J.}~\bibnamefont{Franse}},
  \bibinfo{journal}{Phys.Rev.Lett.} \textbf{\bibinfo{volume}{113}},
  \bibinfo{pages}{251301} (\bibinfo{year}{2014}), \eprint{1402.4119}.

\bibitem[{\citenamefont{Berezinsky and Valle}(1993)}]{Berezinsky:1993fm}
\bibinfo{author}{\bibfnamefont{V.}~\bibnamefont{Berezinsky}} \bibnamefont{and}
  \bibinfo{author}{\bibfnamefont{J.~W.~F.} \bibnamefont{Valle}},
  \bibinfo{journal}{Phys. Lett.} \textbf{\bibinfo{volume}{B318}},
  \bibinfo{pages}{360} (\bibinfo{year}{1993}), \eprint{hep-ph/9309214}.

\bibitem[{\citenamefont{Gu et~al.}(2010)\citenamefont{Gu, Ma, and
  Sarkar}}]{Gu:2010ys}
\bibinfo{author}{\bibfnamefont{P.-H.} \bibnamefont{Gu}},
  \bibinfo{author}{\bibfnamefont{E.}~\bibnamefont{Ma}}, \bibnamefont{and}
  \bibinfo{author}{\bibfnamefont{U.}~\bibnamefont{Sarkar}},
  \bibinfo{journal}{Phys. Lett.} \textbf{\bibinfo{volume}{B690}},
  \bibinfo{pages}{145} (\bibinfo{year}{2010}), \eprint{1004.1919}.

\bibitem[{\citenamefont{Queiroz and Sinha}(2014)}]{Queiroz:2014yna}
\bibinfo{author}{\bibfnamefont{F.~S.} \bibnamefont{Queiroz}} \bibnamefont{and}
  \bibinfo{author}{\bibfnamefont{K.}~\bibnamefont{Sinha}},
  \bibinfo{journal}{Phys. Lett.} \textbf{\bibinfo{volume}{B735}},
  \bibinfo{pages}{69} (\bibinfo{year}{2014}), \eprint{1404.1400}.

\bibitem[{\citenamefont{Lattanzi et~al.}(2014)\citenamefont{Lattanzi, Lineros,
  and Taoso}}]{Lattanzi:2014mia}
\bibinfo{author}{\bibfnamefont{M.}~\bibnamefont{Lattanzi}},
  \bibinfo{author}{\bibfnamefont{R.~A.} \bibnamefont{Lineros}},
  \bibnamefont{and} \bibinfo{author}{\bibfnamefont{M.}~\bibnamefont{Taoso}},
  \bibinfo{journal}{New J. Phys.} \textbf{\bibinfo{volume}{16}},
  \bibinfo{pages}{125012} (\bibinfo{year}{2014}), \eprint{1406.0004}.

\bibitem[{\citenamefont{Boucenna et~al.}(2014)\citenamefont{Boucenna, Morisi,
  Shafi, and Valle}}]{Boucenna:2014uma}
\bibinfo{author}{\bibfnamefont{S.~M.} \bibnamefont{Boucenna}},
  \bibinfo{author}{\bibfnamefont{S.}~\bibnamefont{Morisi}},
  \bibinfo{author}{\bibfnamefont{Q.}~\bibnamefont{Shafi}}, \bibnamefont{and}
  \bibinfo{author}{\bibfnamefont{J.~W.~F.} \bibnamefont{Valle}},
  \bibinfo{journal}{Phys. Rev.} \textbf{\bibinfo{volume}{D90}},
  \bibinfo{pages}{055023} (\bibinfo{year}{2014}), \eprint{1404.3198}.

\bibitem[{\citenamefont{Heeck and Teresi}(2017)}]{Heeck:2017xbu}
\bibinfo{author}{\bibfnamefont{J.}~\bibnamefont{Heeck}} \bibnamefont{and}
  \bibinfo{author}{\bibfnamefont{D.}~\bibnamefont{Teresi}},
  \bibinfo{journal}{Phys. Rev.} \textbf{\bibinfo{volume}{D96}},
  \bibinfo{pages}{035018} (\bibinfo{year}{2017}), \eprint{1706.09909}.

\bibitem[{\citenamefont{Choi and Santamaria}(1990)}]{Choi:1989hi}
\bibinfo{author}{\bibfnamefont{K.}~\bibnamefont{Choi}} \bibnamefont{and}
  \bibinfo{author}{\bibfnamefont{A.}~\bibnamefont{Santamaria}},
  \bibinfo{journal}{Phys. Rev.} \textbf{\bibinfo{volume}{D42}},
  \bibinfo{pages}{293} (\bibinfo{year}{1990}).

\bibitem[{\citenamefont{Pastor et~al.}(1999)\citenamefont{Pastor, Rindani, and
  Valle}}]{Pastor:1997nn}
\bibinfo{author}{\bibfnamefont{S.}~\bibnamefont{Pastor}},
  \bibinfo{author}{\bibfnamefont{S.~D.} \bibnamefont{Rindani}},
  \bibnamefont{and} \bibinfo{author}{\bibfnamefont{J.~W.~F.}
  \bibnamefont{Valle}}, \bibinfo{journal}{JHEP} \textbf{\bibinfo{volume}{05}},
  \bibinfo{pages}{012} (\bibinfo{year}{1999}), \eprint{hep-ph/9705394}.

\bibitem[{\citenamefont{Kachelriess et~al.}(2000)\citenamefont{Kachelriess,
  Tomas, and Valle}}]{Kachelriess:2000qc}
\bibinfo{author}{\bibfnamefont{M.}~\bibnamefont{Kachelriess}},
  \bibinfo{author}{\bibfnamefont{R.}~\bibnamefont{Tomas}}, \bibnamefont{and}
  \bibinfo{author}{\bibfnamefont{J.~W.~F.} \bibnamefont{Valle}},
  \bibinfo{journal}{Phys. Rev.} \textbf{\bibinfo{volume}{D62}},
  \bibinfo{pages}{023004} (\bibinfo{year}{2000}), \eprint{hep-ph/0001039}.

\bibitem[{\citenamefont{Hirsch et~al.}(2009)\citenamefont{Hirsch, Vicente,
  Meyer, and Porod}}]{Hirsch:2009ee}
\bibinfo{author}{\bibfnamefont{M.}~\bibnamefont{Hirsch}},
  \bibinfo{author}{\bibfnamefont{A.}~\bibnamefont{Vicente}},
  \bibinfo{author}{\bibfnamefont{J.}~\bibnamefont{Meyer}}, \bibnamefont{and}
  \bibinfo{author}{\bibfnamefont{W.}~\bibnamefont{Porod}},
  \bibinfo{journal}{Phys. Rev.} \textbf{\bibinfo{volume}{D79}},
  \bibinfo{pages}{055023} (\bibinfo{year}{2009}), \bibinfo{note}{[Erratum:
  Phys. Rev.D79,079901(2009)]}, \eprint{0902.0525}.

\bibitem[{\citenamefont{Garcia~i Tormo et~al.}(2011)\citenamefont{Garcia~i
  Tormo, Bryman, Czarnecki, and Dowling}}]{GarciaiTormo:2011et}
\bibinfo{author}{\bibfnamefont{X.}~\bibnamefont{Garcia~i Tormo}},
  \bibinfo{author}{\bibfnamefont{D.}~\bibnamefont{Bryman}},
  \bibinfo{author}{\bibfnamefont{A.}~\bibnamefont{Czarnecki}},
  \bibnamefont{and} \bibinfo{author}{\bibfnamefont{M.}~\bibnamefont{Dowling}},
  \bibinfo{journal}{Phys. Rev.} \textbf{\bibinfo{volume}{D84}},
  \bibinfo{pages}{113010} (\bibinfo{year}{2011}), \eprint{1110.2874}.

\bibitem[{\citenamefont{Scherrer and Turner}(1985)}]{Scherrer:1984fd}
\bibinfo{author}{\bibfnamefont{R.~J.} \bibnamefont{Scherrer}} \bibnamefont{and}
  \bibinfo{author}{\bibfnamefont{M.~S.} \bibnamefont{Turner}},
  \bibinfo{journal}{Phys. Rev.} \textbf{\bibinfo{volume}{D31}},
  \bibinfo{pages}{681} (\bibinfo{year}{1985}).

\bibitem[{\citenamefont{Bezrukov et~al.}(2010)\citenamefont{Bezrukov,
  Hettmansperger, and Lindner}}]{Bezrukov:2009th}
\bibinfo{author}{\bibfnamefont{F.}~\bibnamefont{Bezrukov}},
  \bibinfo{author}{\bibfnamefont{H.}~\bibnamefont{Hettmansperger}},
  \bibnamefont{and} \bibinfo{author}{\bibfnamefont{M.}~\bibnamefont{Lindner}},
  \bibinfo{journal}{Phys. Rev.} \textbf{\bibinfo{volume}{D81}},
  \bibinfo{pages}{085032} (\bibinfo{year}{2010}), \eprint{0912.4415}.

\bibitem[{\citenamefont{Garcia-Cely et~al.}(2014)\citenamefont{Garcia-Cely,
  Ibarra, and Molinaro}}]{Garcia-Cely:2013wda}
\bibinfo{author}{\bibfnamefont{C.}~\bibnamefont{Garcia-Cely}},
  \bibinfo{author}{\bibfnamefont{A.}~\bibnamefont{Ibarra}}, \bibnamefont{and}
  \bibinfo{author}{\bibfnamefont{E.}~\bibnamefont{Molinaro}},
  \bibinfo{journal}{JCAP} \textbf{\bibinfo{volume}{1402}}, \bibinfo{pages}{032}
  (\bibinfo{year}{2014}), \eprint{1312.3578}.

\bibitem[{\citenamefont{Gu and Sarkar}(2011)}]{Gu:2009hn}
\bibinfo{author}{\bibfnamefont{P.-H.} \bibnamefont{Gu}} \bibnamefont{and}
  \bibinfo{author}{\bibfnamefont{U.}~\bibnamefont{Sarkar}},
  \bibinfo{journal}{Eur. Phys. J.} \textbf{\bibinfo{volume}{C71}},
  \bibinfo{pages}{1560} (\bibinfo{year}{2011}), \eprint{0909.5468}.

\bibitem[{\citenamefont{Chung et~al.}(1999)\citenamefont{Chung, Kolb, and
  Riotto}}]{Chung:1998rq}
\bibinfo{author}{\bibfnamefont{D.~J.~H.} \bibnamefont{Chung}},
  \bibinfo{author}{\bibfnamefont{E.~W.} \bibnamefont{Kolb}}, \bibnamefont{and}
  \bibinfo{author}{\bibfnamefont{A.}~\bibnamefont{Riotto}},
  \bibinfo{journal}{Phys. Rev.} \textbf{\bibinfo{volume}{D60}},
  \bibinfo{pages}{063504} (\bibinfo{year}{1999}), \eprint{hep-ph/9809453}.

\bibitem[{\citenamefont{Hall et~al.}(2010)\citenamefont{Hall, Jedamzik,
  March-Russell, and West}}]{Hall:2009bx}
\bibinfo{author}{\bibfnamefont{L.~J.} \bibnamefont{Hall}},
  \bibinfo{author}{\bibfnamefont{K.}~\bibnamefont{Jedamzik}},
  \bibinfo{author}{\bibfnamefont{J.}~\bibnamefont{March-Russell}},
  \bibnamefont{and} \bibinfo{author}{\bibfnamefont{S.~M.} \bibnamefont{West}},
  \bibinfo{journal}{JHEP} \textbf{\bibinfo{volume}{03}}, \bibinfo{pages}{080}
  (\bibinfo{year}{2010}), \eprint{0911.1120}.

\bibitem[{\citenamefont{Chacko et~al.}(2004)\citenamefont{Chacko, Hall, Okui,
  and Oliver}}]{Chacko:2003dt}
\bibinfo{author}{\bibfnamefont{Z.}~\bibnamefont{Chacko}},
  \bibinfo{author}{\bibfnamefont{L.~J.} \bibnamefont{Hall}},
  \bibinfo{author}{\bibfnamefont{T.}~\bibnamefont{Okui}}, \bibnamefont{and}
  \bibinfo{author}{\bibfnamefont{S.~J.} \bibnamefont{Oliver}},
  \bibinfo{journal}{Phys. Rev.} \textbf{\bibinfo{volume}{D70}},
  \bibinfo{pages}{085008} (\bibinfo{year}{2004}), \eprint{hep-ph/0312267}.

\bibitem[{\citenamefont{Dodelson and Widrow}(1994)}]{Dodelson:1993je}
\bibinfo{author}{\bibfnamefont{S.}~\bibnamefont{Dodelson}} \bibnamefont{and}
  \bibinfo{author}{\bibfnamefont{L.~M.} \bibnamefont{Widrow}},
  \bibinfo{journal}{Phys. Rev. Lett.} \textbf{\bibinfo{volume}{72}},
  \bibinfo{pages}{17} (\bibinfo{year}{1994}), \eprint{hep-ph/9303287}.

\bibitem[{\citenamefont{Boyarsky
  et~al.}(2006{\natexlab{a}})\citenamefont{Boyarsky, Neronov, Ruchayskiy,
  Shaposhnikov, and Tkachev}}]{Boyarsky:2006fg}
\bibinfo{author}{\bibfnamefont{A.}~\bibnamefont{Boyarsky}},
  \bibinfo{author}{\bibfnamefont{A.}~\bibnamefont{Neronov}},
  \bibinfo{author}{\bibfnamefont{O.}~\bibnamefont{Ruchayskiy}},
  \bibinfo{author}{\bibfnamefont{M.}~\bibnamefont{Shaposhnikov}},
  \bibnamefont{and} \bibinfo{author}{\bibfnamefont{I.}~\bibnamefont{Tkachev}},
  \bibinfo{journal}{Phys. Rev. Lett.} \textbf{\bibinfo{volume}{97}},
  \bibinfo{pages}{261302} (\bibinfo{year}{2006}{\natexlab{a}}),
  \eprint{astro-ph/0603660}.

\bibitem[{\citenamefont{Boyarsky et~al.}(2007)\citenamefont{Boyarsky,
  Nevalainen, and Ruchayskiy}}]{Boyarsky:2006ag}
\bibinfo{author}{\bibfnamefont{A.}~\bibnamefont{Boyarsky}},
  \bibinfo{author}{\bibfnamefont{J.}~\bibnamefont{Nevalainen}},
  \bibnamefont{and}
  \bibinfo{author}{\bibfnamefont{O.}~\bibnamefont{Ruchayskiy}},
  \bibinfo{journal}{Astron. Astrophys.} \textbf{\bibinfo{volume}{471}},
  \bibinfo{pages}{51} (\bibinfo{year}{2007}), \eprint{astro-ph/0610961}.

\bibitem[{\citenamefont{Boyarsky
  et~al.}(2006{\natexlab{b}})\citenamefont{Boyarsky, Neronov, Ruchayskiy, and
  Shaposhnikov}}]{Boyarsky:2005us}
\bibinfo{author}{\bibfnamefont{A.}~\bibnamefont{Boyarsky}},
  \bibinfo{author}{\bibfnamefont{A.}~\bibnamefont{Neronov}},
  \bibinfo{author}{\bibfnamefont{O.}~\bibnamefont{Ruchayskiy}},
  \bibnamefont{and}
  \bibinfo{author}{\bibfnamefont{M.}~\bibnamefont{Shaposhnikov}},
  \bibinfo{journal}{Mon. Not. Roy. Astron. Soc.}
  \textbf{\bibinfo{volume}{370}}, \bibinfo{pages}{213}
  (\bibinfo{year}{2006}{\natexlab{b}}), \eprint{astro-ph/0512509}.

\bibitem[{\citenamefont{Boyarsky
  et~al.}(2008{\natexlab{a}})\citenamefont{Boyarsky, Iakubovskyi, Ruchayskiy,
  and Savchenko}}]{Boyarsky:2007ay}
\bibinfo{author}{\bibfnamefont{A.}~\bibnamefont{Boyarsky}},
  \bibinfo{author}{\bibfnamefont{D.}~\bibnamefont{Iakubovskyi}},
  \bibinfo{author}{\bibfnamefont{O.}~\bibnamefont{Ruchayskiy}},
  \bibnamefont{and}
  \bibinfo{author}{\bibfnamefont{V.}~\bibnamefont{Savchenko}},
  \bibinfo{journal}{Mon. Not. Roy. Astron. Soc.}
  \textbf{\bibinfo{volume}{387}}, \bibinfo{pages}{1361}
  (\bibinfo{year}{2008}{\natexlab{a}}), \eprint{0709.2301}.

\bibitem[{\citenamefont{Boyarsky
  et~al.}(2008{\natexlab{b}})\citenamefont{Boyarsky, Malyshev, Neronov, and
  Ruchayskiy}}]{Boyarsky:2007ge}
\bibinfo{author}{\bibfnamefont{A.}~\bibnamefont{Boyarsky}},
  \bibinfo{author}{\bibfnamefont{D.}~\bibnamefont{Malyshev}},
  \bibinfo{author}{\bibfnamefont{A.}~\bibnamefont{Neronov}}, \bibnamefont{and}
  \bibinfo{author}{\bibfnamefont{O.}~\bibnamefont{Ruchayskiy}},
  \bibinfo{journal}{Mon. Not. Roy. Astron. Soc.}
  \textbf{\bibinfo{volume}{387}}, \bibinfo{pages}{1345}
  (\bibinfo{year}{2008}{\natexlab{b}}), \eprint{0710.4922}.

\bibitem[{\citenamefont{Seljak et~al.}(2006)\citenamefont{Seljak, Makarov,
  McDonald, and Trac}}]{Seljak:2006qw}
\bibinfo{author}{\bibfnamefont{U.}~\bibnamefont{Seljak}},
  \bibinfo{author}{\bibfnamefont{A.}~\bibnamefont{Makarov}},
  \bibinfo{author}{\bibfnamefont{P.}~\bibnamefont{McDonald}}, \bibnamefont{and}
  \bibinfo{author}{\bibfnamefont{H.}~\bibnamefont{Trac}},
  \bibinfo{journal}{Phys. Rev. Lett.} \textbf{\bibinfo{volume}{97}},
  \bibinfo{pages}{191303} (\bibinfo{year}{2006}), \eprint{astro-ph/0602430}.

\bibitem[{\citenamefont{Boyarsky et~al.}(2009)\citenamefont{Boyarsky,
  Lesgourgues, Ruchayskiy, and Viel}}]{Boyarsky:2008xj}
\bibinfo{author}{\bibfnamefont{A.}~\bibnamefont{Boyarsky}},
  \bibinfo{author}{\bibfnamefont{J.}~\bibnamefont{Lesgourgues}},
  \bibinfo{author}{\bibfnamefont{O.}~\bibnamefont{Ruchayskiy}},
  \bibnamefont{and} \bibinfo{author}{\bibfnamefont{M.}~\bibnamefont{Viel}},
  \bibinfo{journal}{JCAP} \textbf{\bibinfo{volume}{0905}}, \bibinfo{pages}{012}
  (\bibinfo{year}{2009}), \eprint{0812.0010}.

\bibitem[{\citenamefont{Horiuchi et~al.}(2014)\citenamefont{Horiuchi, Humphrey,
  Onorbe, Abazajian, Kaplinghat, and Garrison-Kimmel}}]{Horiuchi:2013noa}
\bibinfo{author}{\bibfnamefont{S.}~\bibnamefont{Horiuchi}},
  \bibinfo{author}{\bibfnamefont{P.~J.} \bibnamefont{Humphrey}},
  \bibinfo{author}{\bibfnamefont{J.}~\bibnamefont{Onorbe}},
  \bibinfo{author}{\bibfnamefont{K.~N.} \bibnamefont{Abazajian}},
  \bibinfo{author}{\bibfnamefont{M.}~\bibnamefont{Kaplinghat}},
  \bibnamefont{and}
  \bibinfo{author}{\bibfnamefont{S.}~\bibnamefont{Garrison-Kimmel}},
  \bibinfo{journal}{Phys. Rev.} \textbf{\bibinfo{volume}{D89}},
  \bibinfo{pages}{025017} (\bibinfo{year}{2014}), \eprint{1311.0282}.

\bibitem[{\citenamefont{Shakya}(2016)}]{Shakya:2015xnx}
\bibinfo{author}{\bibfnamefont{B.}~\bibnamefont{Shakya}},
  \bibinfo{journal}{Mod. Phys. Lett.} \textbf{\bibinfo{volume}{A31}},
  \bibinfo{pages}{1630005} (\bibinfo{year}{2016}), \eprint{1512.02751}.

\bibitem[{\citenamefont{Shakya and Wells}(2017)}]{Shakya:2016oxf}
\bibinfo{author}{\bibfnamefont{B.}~\bibnamefont{Shakya}} \bibnamefont{and}
  \bibinfo{author}{\bibfnamefont{J.~D.} \bibnamefont{Wells}},
  \bibinfo{journal}{Phys. Rev.} \textbf{\bibinfo{volume}{D96}},
  \bibinfo{pages}{031702} (\bibinfo{year}{2017}), \eprint{1611.01517}.

\bibitem[{\citenamefont{Merle et~al.}(2014)\citenamefont{Merle, Niro, and
  Schmidt}}]{Merle:2013wta}
\bibinfo{author}{\bibfnamefont{A.}~\bibnamefont{Merle}},
  \bibinfo{author}{\bibfnamefont{V.}~\bibnamefont{Niro}}, \bibnamefont{and}
  \bibinfo{author}{\bibfnamefont{D.}~\bibnamefont{Schmidt}},
  \bibinfo{journal}{JCAP} \textbf{\bibinfo{volume}{1403}}, \bibinfo{pages}{028}
  (\bibinfo{year}{2014}), \eprint{1306.3996}.

\bibitem[{\citenamefont{Adulpravitchai and
  Schmidt}(2015)}]{Adulpravitchai:2014xna}
\bibinfo{author}{\bibfnamefont{A.}~\bibnamefont{Adulpravitchai}}
  \bibnamefont{and} \bibinfo{author}{\bibfnamefont{M.~A.}
  \bibnamefont{Schmidt}}, \bibinfo{journal}{JHEP}
  \textbf{\bibinfo{volume}{01}}, \bibinfo{pages}{006} (\bibinfo{year}{2015}),
  \eprint{1409.4330}.

\bibitem[{\citenamefont{Kang}(2015)}]{Kang:2014cia}
\bibinfo{author}{\bibfnamefont{Z.}~\bibnamefont{Kang}}, \bibinfo{journal}{Eur.
  Phys. J.} \textbf{\bibinfo{volume}{C75}}, \bibinfo{pages}{471}
  (\bibinfo{year}{2015}), \eprint{1411.2773}.

\bibitem[{\citenamefont{Merle and Totzauer}(2015)}]{Merle:2015oja}
\bibinfo{author}{\bibfnamefont{A.}~\bibnamefont{Merle}} \bibnamefont{and}
  \bibinfo{author}{\bibfnamefont{M.}~\bibnamefont{Totzauer}},
  \bibinfo{journal}{JCAP} \textbf{\bibinfo{volume}{1506}}, \bibinfo{pages}{011}
  (\bibinfo{year}{2015}), \eprint{1502.01011}.

\bibitem[{\citenamefont{Gorbunov and Shaposhnikov}(2007)}]{Gorbunov:2007ak}
\bibinfo{author}{\bibfnamefont{D.}~\bibnamefont{Gorbunov}} \bibnamefont{and}
  \bibinfo{author}{\bibfnamefont{M.}~\bibnamefont{Shaposhnikov}},
  \bibinfo{journal}{JHEP} \textbf{\bibinfo{volume}{10}}, \bibinfo{pages}{015}
  (\bibinfo{year}{2007}), \bibinfo{note}{[Erratum: JHEP11,101(2013)]},
  \eprint{0705.1729}.

\bibitem[{\citenamefont{Adams et~al.}(2013)}]{Adams:2013qkq}
\bibinfo{author}{\bibfnamefont{C.}~\bibnamefont{Adams}} \bibnamefont{et~al.}
  (\bibinfo{collaboration}{LBNE}) (\bibinfo{year}{2013}), \eprint{1307.7335}.

\bibitem[{\citenamefont{Jacobsson}(2016)}]{Jacobsson:2130433}
\bibinfo{author}{\bibfnamefont{R.}~\bibnamefont{Jacobsson}}
  (\bibinfo{collaboration}{SHiP Collaboration}), \bibinfo{type}{Tech. Rep.}
  \bibinfo{number}{CERN-SHiP-PROC-2016-007}, \bibinfo{institution}{CERN},
  \bibinfo{address}{Geneva} (\bibinfo{year}{2016}),
  \urlprefix\url{https://cds.cern.ch/record/2130433}.

\bibitem[{\citenamefont{Kusenko}(2009)}]{Kusenko:2009up}
\bibinfo{author}{\bibfnamefont{A.}~\bibnamefont{Kusenko}},
  \bibinfo{journal}{Phys. Rept.} \textbf{\bibinfo{volume}{481}},
  \bibinfo{pages}{1} (\bibinfo{year}{2009}), \eprint{0906.2968}.

\bibitem[{\citenamefont{Hernandez et~al.}(2014)\citenamefont{Hernandez, Kekic,
  and Lopez-Pavon}}]{Hernandez:2014fha}
\bibinfo{author}{\bibfnamefont{P.}~\bibnamefont{Hernandez}},
  \bibinfo{author}{\bibfnamefont{M.}~\bibnamefont{Kekic}}, \bibnamefont{and}
  \bibinfo{author}{\bibfnamefont{J.}~\bibnamefont{Lopez-Pavon}},
  \bibinfo{journal}{Phys. Rev.} \textbf{\bibinfo{volume}{D90}},
  \bibinfo{pages}{065033} (\bibinfo{year}{2014}), \eprint{1406.2961}.

\bibitem[{\citenamefont{Vincent et~al.}(2015)\citenamefont{Vincent, Martinez,
  Hernández, Lattanzi, and Mena}}]{Vincent:2014rja}
\bibinfo{author}{\bibfnamefont{A.~C.} \bibnamefont{Vincent}},
  \bibinfo{author}{\bibfnamefont{E.~F.} \bibnamefont{Martinez}},
  \bibinfo{author}{\bibfnamefont{P.}~\bibnamefont{Hernández}},
  \bibinfo{author}{\bibfnamefont{M.}~\bibnamefont{Lattanzi}}, \bibnamefont{and}
  \bibinfo{author}{\bibfnamefont{O.}~\bibnamefont{Mena}},
  \bibinfo{journal}{JCAP} \textbf{\bibinfo{volume}{1504}}, \bibinfo{pages}{006}
  (\bibinfo{year}{2015}), \eprint{1408.1956}.

\bibitem[{\citenamefont{Chikashige et~al.}(1980)\citenamefont{Chikashige,
  Mohapatra, and Peccei}}]{Chikashige:1980qk}
\bibinfo{author}{\bibfnamefont{Y.}~\bibnamefont{Chikashige}},
  \bibinfo{author}{\bibfnamefont{R.~N.} \bibnamefont{Mohapatra}},
  \bibnamefont{and} \bibinfo{author}{\bibfnamefont{R.~D.}
  \bibnamefont{Peccei}}, \bibinfo{journal}{Phys. Rev. Lett.}
  \textbf{\bibinfo{volume}{45}}, \bibinfo{pages}{1926} (\bibinfo{year}{1980}).

\bibitem[{\citenamefont{Garcia-Cely et~al.}(2013)\citenamefont{Garcia-Cely,
  Ibarra, and Molinaro}}]{Garcia-Cely:2013nin}
\bibinfo{author}{\bibfnamefont{C.}~\bibnamefont{Garcia-Cely}},
  \bibinfo{author}{\bibfnamefont{A.}~\bibnamefont{Ibarra}}, \bibnamefont{and}
  \bibinfo{author}{\bibfnamefont{E.}~\bibnamefont{Molinaro}},
  \bibinfo{journal}{JCAP} \textbf{\bibinfo{volume}{1311}}, \bibinfo{pages}{061}
  (\bibinfo{year}{2013}), \eprint{1310.6256}.

\bibitem[{\citenamefont{Weinberg}(2013)}]{Weinberg:2013kea}
\bibinfo{author}{\bibfnamefont{S.}~\bibnamefont{Weinberg}},
  \bibinfo{journal}{Phys. Rev. Lett.} \textbf{\bibinfo{volume}{110}},
  \bibinfo{pages}{241301} (\bibinfo{year}{2013}), \eprint{1305.1971}.

\bibitem[{\citenamefont{Josse-Michaux and
  Molinaro}(2011)}]{JosseMichaux:2011ba}
\bibinfo{author}{\bibfnamefont{F.-X.} \bibnamefont{Josse-Michaux}}
  \bibnamefont{and} \bibinfo{author}{\bibfnamefont{E.}~\bibnamefont{Molinaro}},
  \bibinfo{journal}{Phys. Rev.} \textbf{\bibinfo{volume}{D84}},
  \bibinfo{pages}{125021} (\bibinfo{year}{2011}), \eprint{1108.0482}.

\end{thebibliography}

\end{document}